\documentclass[%
 reprint,
superscriptaddress,
nofootinbib,
 amsmath,amssymb,
 aps,
 prapplied
]{revtex4-2}

\usepackage{graphicx}
\usepackage{physics}
\usepackage{dcolumn}
\usepackage{tikz}
\usepackage{pgfplots}
\usepackage{bm}
\usepackage{dsfont}
\usepackage{soul}
\pgfplotsset{compat=1.6}
\usepackage{mathtools}
\usepackage{array}
\usepackage{pgfplots}
\usepackage{pgfplotstable}
\usepackage[capitalize]{cleveref}

\definecolor{color1bg}{HTML}{1f77b4}
\definecolor{color2bg}{HTML}{ff7f0e}
\definecolor{color3bg}{HTML}{2ca02c}
\definecolor{color4bg}{HTML}{d62728}
\definecolor{color5bg}{HTML}{9467bd}
\definecolor{color6bg}{HTML}{8c564b}
\definecolor{colorfill}{HTML}{fbf19a}

\newcolumntype{M}[1]{>{\centering\arraybackslash}m{#1}}

\pgfplotsset{
  /pgfplots/error bar legend/.style={
    legend image code/.code={
        \draw[sharp plot,mark=-,mark repeat=2,mark size = 4 pt, line width = 2pt, mark phase=1,color=#1]
        plot coordinates { (0.3cm, -0.15cm) (0.3cm,0cm) (0.3cm, 0.15cm) };%
        
}}}

\makeatletter
\newcounter{subfigure}

\makeatother

\tikzstyle{startstop} = [rectangle, rounded corners, minimum width=3cm, minimum height=1cm,
    text width=5cm,  align=center, draw=black, fill=red!30]

\tikzstyle{process} = [rectangle, minimum width=3cm, minimum height=1cm,
    text width=6cm, align=center, draw=black, fill=orange!30]

\tikzstyle{thickprocess} = [rectangle, minimum width=8cm, minimum height=2cm,
    text width=8cm, align=center, draw=black, fill=orange!30]
\tikzstyle{arrow} = [thick,->,>=stealth]

\pgfplotsset{
  /pgfplots/error bar legend/.style={
    legend image code/.code={
        \draw[sharp plot,mark=-,mark repeat=2,mark size = 4 pt, line width = 2pt, mark phase=1,color=#1]
        plot coordinates { (0.3cm, -0.15cm) (0.3cm,0cm) (0.3cm, 0.15cm) };%
        
}}}

\begin{document}

\preprint{APS/123-QED}

\title{Random Party Distillation on a Superconducting Processor}

\author{Alexander C.B. Greenwood}
\email{alexander.greenwood@mail.utoronto.ca}

\author{Jackson Russett}
\affiliation{Dept of Electrical \& Computer Engineering, University of Toronto \\
Toronto, Ontario, Canada M5S 3G4}
\author{Hoi-Kwong Lo}
\affiliation{Dept of Electrical \& Computer Engineering, University of Toronto \\
Toronto, Ontario, Canada M5S 3G4}
\affiliation{Department of Physics,
National University of Singapore
Blk S12 Level 2, 2 Science Drive 3
Singapore 117551}
\affiliation{Quantum Bridge Technologies Inc., 108 College Street, Toronto, Canada}
\affiliation{Centre for Quantum Technologies, National University of Singapore, Singapore 117543}
\author{Li Qian}

\affiliation{Dept of Electrical \& Computer Engineering, University of Toronto \\
Toronto, Ontario, Canada M5S 3G4}

\date{\today}

\begin{abstract}
Random party distillation refers to the process by which Einstein-Podolsky-Rosen pairs are probabilistically extracted from a single copy of a multipartite entangled state after multiple rounds of local operations and classical communications (LOCC). In this work, we propose a qubit-based implementation of a random party distillation protocol and demonstrate it on the superconducting hardware device, \texttt{ibm\_aachen}. We implement up to 4-rounds of the protocol, achieving a distillation rate of $\sim0.85$ pairs/ \textit{W} state, surpassing previous experimental demonstrations. Our results demonstrate the utility of multi-round LOCC protocols in an experimental regime that remains comparatively unexplored.


\end{abstract}

\maketitle




\section{Introduction}
Entanglement is the essence of quantum mechanics \cite{QuantEntangle}.
The task of establishing bipartite correlations between physically distant parties is a basic subroutine necessary for quantum communications, particularly while teleporting qubits or gate operations. An active area of research is concerned with the distribution of entanglement between laboratories separated by hundreds of kilometers of optical fiber \cite{sun2017entanglement}, or transmon qubits that are spaced hundreds of micrometers to millimeters apart \cite{PRXQuantum.5.030339, PhysRevApplied.23.014057}.

Random party distillation \cite{PhysRevLett.98.260501}, is an interesting paradigm for studying multipartite entanglement and its conversion to bipartite entanglement. Unlike the bipartite case, multipartite entanglement lacks consensus of a ``maximally-entangled" state, which depends on one's choice of measure \cite{enriquez2016maximally}. For three qubits, this richness is already apparent from the existence of two inequivalent classes of genuine tripartite entanglement: the GHZ and W classes \cite{PhysRevA.62.062314,PhysRevLett.87.040401}.
One can, of course, quantify entanglement operationally, through a state’s ability to produce bipartite entanglement. Examples include concentratable entanglement \cite{PhysRevLett.127.140501}, entanglement of assistance \cite{PhysRevA.72.042318,PhysRevA.111.032423}, and the monotones in Refs. \cite{PhysRevA.85.062316,PhysRevLett.108.240504}. The last of these characterizes a state’s usefulness for random-party distillation, in which a single copy of a \textit{W}-like state is subjected to repeated measurements and, with some probability, yields an Einstein–Podolsky–Rosen (EPR) pair after up to N rounds of the protocol. The W state is especially natural in this setting because its entanglement persists under particle loss and can be probabilistically localized into bipartite entanglement by measuring one party.
%

With random party distillation, The probability of obtaining a single EPR pair from one copy of a W state converges to unity in the limit of many rounds $N\to\infty$ \cite{PhysRevLett.98.260501}. The main benefit of random party distillation is that it enables the distribution of bipartite entanglement to parties possessing a pre-shared multipartite entangled state, relying on local operations and classical communication (LOCC) alone. And thus, random party distillation is useful in scenarios where multipartite entangled states may be directly initialized on a processor, such as in Ref. \cite{roy2025parity}, or via interactions between stationary and multi-qubit states encoded in flying qubits (such as photons) \cite{PRXQuantum.5.010202}. Indeed, the latter scenario is potentially impactful for distributed quantum computing scenarios on either superconducting, trapped-ion, and neutral atom platforms \cite{PRXQuantum.5.010202}.

Despite its proposed advantages, random party distillation has been studied mainly in theory, including in the first demonstration of a gap between separable (SEP) operations and LOCC \cite{PhysRevLett.108.240504}, analyses of the round complexity of LOCC \cite{Liu2023roundcomplexity}, and proposals for improving resolution limits in long-baseline telescopy \cite{wang2024randomdistillationprotocolslong}. By contrast, experimental realizations have so far been limited to single-round implementations \cite{OpticalSagnacRPD}, which restrict the achievable success probability. More generally, experimental demonstrations of multi-round LOCC protocols remain rare, with notable exceptions in Refs. \cite{PhysRevLett.105.230502,abdelkhalek2016efficient}. A multi-round implementation of random party distillation could therefore both enhance distillation performance in practice and highlight the utility of a regime of LOCC that has remained largely unexplored experimentally.

In this article, we present an experimental demonstration of random party distillation on the superconducting processor, \texttt{ibm\_aachen}, that is capable of executing over four rounds of the protocol with a distillation rate of 0.85 pairs/ \textit{W} state (superior to existing implementations). We introduce a generalized version of the protocol originally proposed by Fortescue and Lo \cite{PhysRevLett.98.260501}, adapted for qubit systems without requiring access to higher energy levels. Creating multiparticle entanglement with transmon qubits and shielding it from dephasing is not an easy task; we show how the number of rounds of execution, and thus, success probability of our protocol, can be extended by means of dynamical decoupling and matrix-free measurement error mitigation (M3) \cite{PRXQuantum.2.040326}. 




\section{EPR Distillation from Multipartite States}
We consider EPR distillation with three parties, although our method can be extended to larger multiparticle systems, using protocols described in \cite{PhysRevA.78.012348, PhysRevA.84.052301}. Alice, Bob, and Charlie wish to distill the triplet state:
\begin{align}
    \ket{\Psi^+}_{ij} = &\frac{1}{\sqrt{2}}\left( \ket{01} + \ket{10} \right)_{ij} \nonumber \\ 
    &\text{ where } i,j \in \{a,b,c\} \text{ and } i \neq j,
    \label{eqn:triplet-state}
\end{align}
from a \textit{W}-class state of the form

\begin{equation}
    \ket{W}_{abc} = \frac{1}{\sqrt{3}}\left( \ket{011} + \ket{101} + \ket{110} \right)_{abc}.
    \label{eqn:w-state}
\end{equation}

The simplest way for Alice, Bob, and Charlie to proceed is to specify one party to perform a strong measurement on their respective qubit in the Z basis. If the specified party measures a ``1'', the other two parties will share an EPR pair. Measurement of a ``0'' indicates that the other parties share a product state. In this protocol, called \textit{specific party distillation}, the probability of successful distillation $P(EPR) = 2/3$, from a single copy of a \textit{W} state according to \eqref{eqn:w-state}. If we wanted to increase the probability of a successful distillation event, we would need to use a method that better leverages the shared entanglement of all three parties.

By allowing the randomness of which pair of parties ultimately share an EPR pair, one can increase the success rate. This \textit{random party distillation} protocol, proposed by Fortescue and Lo \cite{PhysRevLett.98.260501}, consists of each party coupling their qubit to a higher-dimensional Hilbert space, followed by a series of projective measurements to determine the presence of two-qubit entanglement while minimizing disturbance to the system itself. Mathematically, the measurement step at each round of protocol execution can be fully described by a positive operator-valued measure (POVM). Later implementations such as \cite{Liu2023roundcomplexity} or \cite{OpticalSagnacRPD} generalize the action of coupling to a higher-dimensional Hilbert space to a weak measurement process. Unlike projective (strong) measurements that completely destroy entanglement, weak measurements allow subsystems to remain entangled, albeit with less information extraction. 

The objective of random party distillation is to increase the probability of extracting an EPR pair $P(EPR)$ to unity for a \textit{single} copy of the W state in the asymptotic limit of LOCC operation rounds. This is a class of protocol where the probability of success is defined for a single copy of a \textit{W} state reused over multiple rounds of protocol execution. It is important to note the contrast with ``entanglement of assistance" \cite{PhysRevA.72.042318}, where one repeats a protocol \textit{many} identically prepared copies of a state until success.

While the original formulation of the random distillation considers a distant-party setting where the three parties, Alice, Bob and Charles are far away from each other, the essence of the protocol can be demonstrated on a local quantum computer. In what follows, we describe the random distillation protocol on the superconducting hardware device, \texttt{ibm\_aachen}. We detail the implementation of our random distillation protocol on \texttt{ibm\_aachen} superconducting hardware and demonstrate how weak measurements can be realized on such hardware by introducing ancillary qubits coupled to the system of target qubits. 

\begin{figure*}
    \centering
    \includegraphics[width=0.9\linewidth]{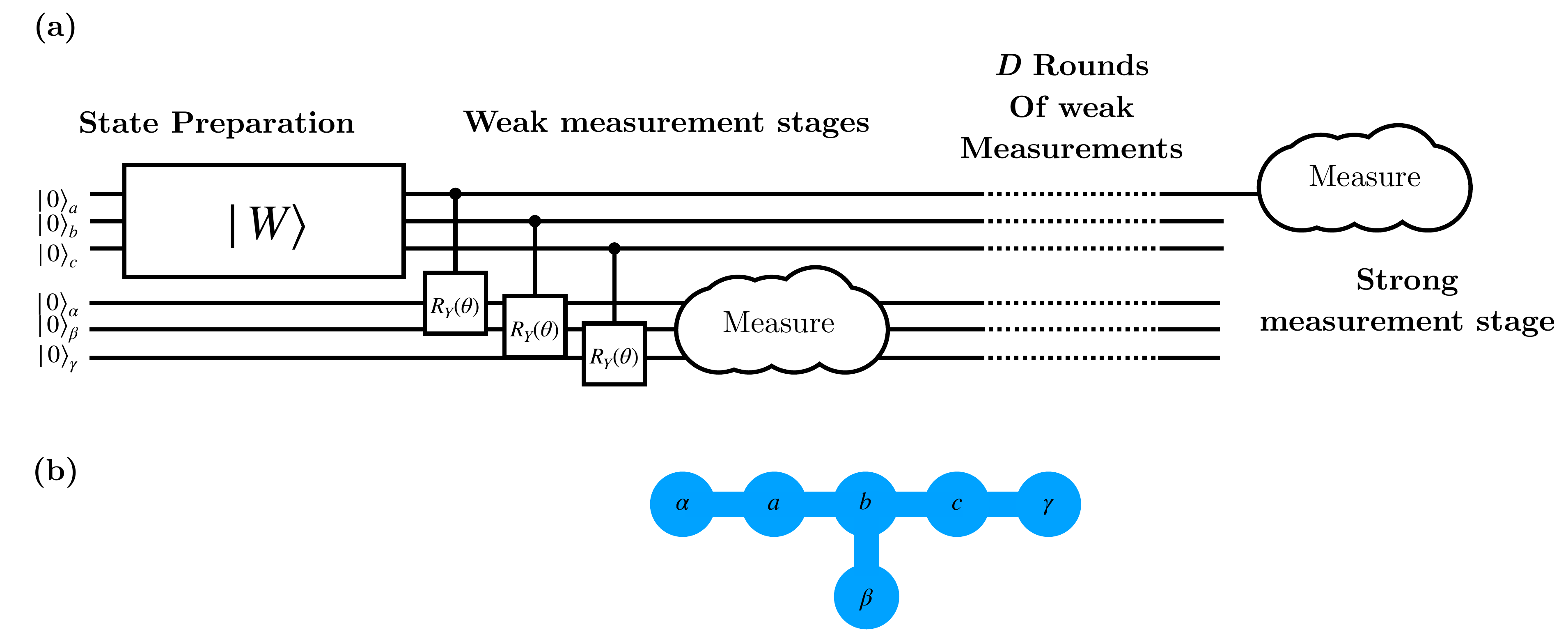}
    \caption{(a) The circuit used to implement a D-round random party distillation protocol. Each controlled $R_y(\theta)$ operation facilitates the coupling between a single party's qubit and its respective ancilla. A measurement on each parties ancillary qubit effectively corresponds to a weak measurement of the system qubits: $a, b, $ and $c$. If after $D$ rounds of measurements the protocol has not succeeded, one of the three parties (Alice) will perform a strong measurement on their qubit. (b) An illustration of the qubit connectivity required to implement the circuit in (a).}
    \label{fig:rand_party_distillation}
\end{figure*}



Suppose Alice, Bob, and Charlie start with a W-state of the form in \eqref{eqn:w-state}. They apply the following operation to the state, coupling their respective qubits to the ancillae with an $ 0 < \epsilon \ll 1$:

\begin{align}
    \ket{0}_{A_i}\ket{1}_i &\to \sqrt{1-\epsilon^2}\ket{0}_{A_i}\ket{1}_i + \epsilon\ket{1}_{A_i}\ket{1}_i,\nonumber\\
    \ket{0}_{A_i}\ket{0}_i &\to \ket{0}_{A_i}\ket{0}_i,
    \label{eq:weak_transform}
\end{align}
where \textit{i} denotes the qubit in the \textit{W} state (\textit{a,b,c}) and $A_i$ is for the corresponding ancilla ($\alpha, \beta, \gamma$).  Each ancilla starts in its ground state (in the Z-basis), $\ket{0}$, and $\epsilon$ represents the strength of the coupling measurement. After this operation, the parties possess the following state:


\begin{align}
    \ket{W'} = &\left(1-\epsilon^2\right)\ket{000}_{\alpha\beta\gamma}\ket{W}_{abc} \nonumber \\
    + &\epsilon \sqrt{\frac{2(1-\epsilon^2)}{3}} \left[\ket{001}_{\alpha\beta\gamma}\ket{1}_c\ket{\Psi^+}_{ab} \nonumber \right. \\ 
    + &\ket{010}_{\alpha\beta\gamma}\ket{1}_b\ket{\Psi^+}_{ac} \nonumber 
     + \left.\ket{100}_{\alpha\beta\gamma}\ket{1}_a\ket{\Psi^+}_{bc} \right] \nonumber \\
    + &O(\epsilon^2).
    \label{eq:state_after_coupling}
\end{align}

The three parties now each make a Z-basis measurement on their respective ancilla using the following projectors:
$F = \ket{0}_{A_i}\bra{0}_{A_i}$ and $G = \ket{1}_{A_i} \bra{1}_{A_i}$.
We can identify the possible outcomes of the three measurements based on the terms in \eqref{eq:state_after_coupling}. Measuring ``0" corresponds to F and ``1" to G. If two parties obtain an outcome of ``0" and one party obtains an outcome of ``1", then the two parties that measured ``0" possess an EPR pair among themselves. If two or more parties see an outcome of ``1" then protocol ends immediately without distillation; the protocol fails. If all three parties obtain an outcome of ``0", then they possess a \textit{W} state and the protocol repeats itself. All relevant results from the measurements of the ancillae are summarized in Table \ref{tab:transform_outcome_prob}.

After all $N$ rounds are complete, an additional strong measurement step is added, unlike the original proposal in \cite{QutritRPDTheory}. This is shown in Figure \ref{fig:rand_party_distillation} where the final strong measurement is made on Alice's qubit. This means that when all rounds of measurement end in a W-state, a final attempt remains with a 2/3 probability of successfully obtaining an EPR state between Bob and Charlie. This additional step, as proposed in \cite{OpticalSagnacRPD}, helps increase the overall success probability when only a few rounds of measurements are accessible.

\renewcommand{\arraystretch}{2}
\begin{table}
    \centering
    \begin{tabular}{p{7em}p{6em}p{5.5em}p{5.5em}} \hline \hline 
         \textbf{Measurement outcome}& \textbf{Probability}&  \textbf{Resulting state}& \textbf{Next step}\\ \hline 
         $\ket{000}_{\alpha\beta\gamma}$&  $(1-\epsilon^2)^2$&  $\ket{W}_{abc}$& repeat \\ 
         $\ket{100}_{\alpha\beta\gamma}$  & $\frac{2\epsilon^2(1-\epsilon^2)}{3}$ &  $\ket{\Psi^+}_{bc}$ & exit - success \\  $\ket{010}_{\alpha\beta\gamma}$ & $\frac{2\epsilon^2(1-\epsilon^2)}{3}$ & $\ket{\Psi^+}_{ac}$ & exit - success \\   $\ket{001}_{\alpha\beta\gamma}$&  $\frac{2\epsilon^2(1-\epsilon^2)}{3}$ & $\ket{\Psi^+}_{ab}$& exit - success\\  
         ... &  $\epsilon^4$&  N/A & exit - failure\\ \hline
    \end{tabular}
    \caption{Alice, Bob, and Charlie's outcomes after each stage of weak measurements, and their respective probabilities of occurrence.}
    \label{tab:transform_outcome_prob}
\end{table}

Weak measurements are performed on superconducting qubits using a controlled rotation gate, $CRY(\theta)$. Each round of the protocol consists of Alice, Bob, and Charlie applying the controlled rotation between their respective qubits and ancillae. A matrix representation of the $CRY(\theta)$ operation is given as follows: 

\begin{equation}
    CRY(\theta)_{A_i,i} = \mathds{1}\otimes\ket{0}\!\!\bra{0} + RY(\theta)\otimes \ket{1}\!\!\bra{1} 
\end{equation}
The choice of $\theta$ is related to the measurement strength $\epsilon$ in Eq. (\ref{eq:weak_transform}) by $\theta  = 2\arcsin\left(\epsilon\right)$.


For simplicity, we assume that, in each round of the protocol, all three parties utilize the same value $\epsilon$ (which can be round-dependent). Thus, the same measurement strength, for their operations and choose $\epsilon$ to maximize the probability of eventually distilling an EPR pair. For $N$ rounds of measurements, the maximum success probability is

\begin{equation}
    P_{N} = \frac{N+2}{N+3},
    \label{eq:success_prob_theory}
\end{equation}

when $\epsilon$ is optimized for each round as proposed in \cite{OpticalSagnacRPD}:

\begin{equation}
    \epsilon_D = \frac{1}{\sqrt{D+5}}.
    \label{eq:opt_epsilon}
\end{equation}

The value of $\epsilon_D$ is determined by the number of \textit{remaining} rounds, $D$, in the protocol. This means that each measurement round is stronger than the preceding round. Of course, Eqs. \eqref{eq:success_prob_theory} and \eqref{eq:opt_epsilon} assume that we began with an ideal \textit{W} state, which is never the case on NISQ (near intermediate scale quantum) hardware. To this end, we shall now describe the characterization of the physical implementation and compare its performance to the theory.


Our random party distillation experiment was carried out on the 156-qubit Heron r3 system, \texttt{ibm\_aachen}. The protocol outlined in this work is applicable to modalities of quantum computing outside of transmon qubits, e.g. trapped-ion, photonic qubits, etc.
Fig. \ref{fig:rand_party_distillation}b shows an example of a connectivity graph between qubits on \texttt{ibm\_aachen}, where edges between qubits indicate that an entanglement gate operation (in this case, an echoed cross-resonance gate) may be performed and the nodes represent qubits. 
While we consider the archetypal example of a 3-qubit W state for extracting single pairs of EPR states in this work, one could also consider larger systems larger than three qubits for extracting arbitrary configurations of bipartite (and multipartite) entanglement using the result from Ref. \cite{PhysRevA.84.052301}.

The protocol outlined in Fig. \ref{fig:rand_party_distillation} requires \textit{dynamic circuits}: circuits whose operations are determined by the results of its mid-circuit measurements \cite{PRXQuantum.5.030339}.
In this case, the protocol only continues if Alice, Bob, and Charlie simultaneously measure a ``0" bit during a round of weak mid-circuit measurements. This outcome indicates that a W-state is still present and that another round of weak measurements can be performed. The circuit depth for our protocol thus ranges from ~24 (counting basis gates and measurements) for a single-round implementation of the protocol to up to ~60 for a four-round protocol (as shown in this work). We find that each round of protocol execution on average requires 12 operations. 

The following characterization steps: initial W-state fidelity, measurement of distillation rate, evolution of state fidelity, and entanglement of formation have all been considered using static circuits with mid-circuit measurement. 
Functionally, these circuits are nearly equivalent to the dynamic case, except that the success of distillation is determined after executing a predetermined number of rounds of execution rather than completing at the first detection of bipartite entanglement. 
An experimental challenge with mid-circuit measurements in IBM quantum computers is that the measurement time is rather long ($\sim$1$\mu$s), during which idle qubits may decohere. To reduce errors due to decoherence during the mid-circuit
measurement, it is important to apply error mitigation techniques such as dynamical
decoupling. For information pertaining to the coherence and dephasing time of qubits used in this work, the reader is encouraged to consult the corresponding supplementary material.



After each round of Alice, Bob, and Charlie applying $CRY(\theta)$ operations, the three parties make Z-basis measurements on their ancilla qubits. The probability of the protocol succeeding, two parties sharing an EPR pair can be expressed as

\begin{multline}
    P_{success} = P_1(EPR) + \sum_{k=2}^{N} P_{k-1}(W)P_k(EPR|W) \\+ P_s(EPR|W_N)P_N(W),
    \label{eq:bayes_success_probability}
\end{multline}

where $P_k(EPR)$ and $P_k(W)$ correspond to the probabilities of obtaining an EPR pair and W state, respectively, in round $k$ out of a total of $N$ rounds. $P_s(EPR|W_N)$ corresponds to the probability of obtaining an EPR during the strong measurement stage given that a \textit{W} state was detected during the final stage of weak measurements. These probabilities can be experimentally determined by preparing many copies of a \textit{W} state, and repeating the execution of the protocol. By counting the number of instances in which all three parties measure ``0" in a given round $k$, one can determine the probability of obtaining a \textit{W} state. Likewise, by counting the number of instances in which one of the three parties measures ``1", the probability of successfully distilling an EPR pair may be obtained. 


So far, we have considered an ideal implementation of random party distillation with perfect qubits and gate operations. Now we consider an imperfect scenario where the \textit{W} state that we began with at each round of execution is not ideal and gate operations performed by Alice, Bob, and Charlie have non-unity fidelity.
Consequently, the \textit{expected} amount of bipartite entanglement $\expval{E}$ distilled from the protocol should be calculated by weighting each term in \eqref{eq:bayes_success_probability} by an entanglement measure $E(\rho_{i}^k)$ of the corresponding distilled pair:  

\begin{multline}
    \expval{E} = \frac{1}{|\mathcal{S}|}\sum_{i\in\mathcal{S}}P_1(EPR)E(\rho_{i}^1) 
    \\ + \frac{1}{|\mathcal{S}|}\sum_{i\in\mathcal{S}}\sum_{k=2}^{N} P_{k-1}(W)P_k(EPR|W)E(\rho_{i}^k) \\+ P_s(EPR|W_N)P_N(W)E(\rho_{a}^s)
    \label{eq:weighted_success_probability}
\end{multline}

where $\rho_i^k$ is the bipartite state extracted in round $k$ when the $i$th party measures an outcome of $G$ during their respective projective measurement stage. We have denoted $\rho_a^s$ as the distilled pair possessed by Bob and Charlie after successfully completing the final strong measurement stage. The set $\mathcal{S}$ is defined as the set of parties involved in the protocol. For the purpose of this letter, $\mathcal{S} = \{a,b,c\}$, but one can also consider distillation from $W$ states consisting of more than three parties \cite{PhysRevA.78.012348}. Although previous studies  \cite{OpticalSagnacRPD} report a success probability alone, we make the important distinction between success probability and the expected entanglement. The latter quantity provides a realistic measure of the true utility of a random party distillation protocol for the purpose of entanglement distribution. In subsequent sections, we define our measure $E(\rho_{i}^k)$ as the entanglement of formation according to \cite{PhysRevLett.80.2245}. For a pure state $\ket{\psi_{ij}}$, this results in:

\begin{equation}
    E(\ket{\psi_{ij}}) = -\Tr{\rho_i \log_2 \rho_i} = -\Tr{\rho_j \log_2 \rho_j},
\end{equation}

where $\rho_i$ corresponds to one subsystem when the other qubit has been traced over: 

\begin{equation}
    \rho_i = \Tr_j \ket{\psi_{ij}}\!\!\bra{\psi_{ij}}.
\end{equation}

For a mixed state $\rho$, this simply becomes  the \textit{minimum} average entanglement of formation over all decompositions of $\rho$ \cite{PhysRevLett.80.2245}:

\begin{equation}
    E(\rho) = \min \sum_k p_k E(\ket{\psi^k_{ij}}).
    \label{eq:mixed_eof}
\end{equation}

At first glance, \eqref{eq:mixed_eof} is an unwiedily calculated calculation; however, it can be greatly simplified by obtaining from a closed-form expression dependent on the Wooters concurrence \cite{PhysRevLett.80.2245} or bounding with a quantitative witness described in \cite{eisert2007quantitative}. 

Replacing the measure $E(\rho_{i}^k)$ with \textit{distillable entanglement} would indeed be a natural choice, however, we opt to use entanglement of formation for ease of finding an experimental lower bound using either linear witnesses \cite{eisert2007quantitative} or state fidelity, as will be shown later\footnote{It is possible, however, to obtain a lower bound for distillable entanglement using the entropic uncertainty relations described in \cite{PhysRevLett.126.190503} or from stabilizer measurements, shown in \cite{lo2001proof}. These limits have not been proved to be tight.}.
Furthermore, the entanglement of formation is known to serve as an upper bound to the distillable entanglement \cite{PhysRevA.54.3824}.

\section{Physical Experiment}

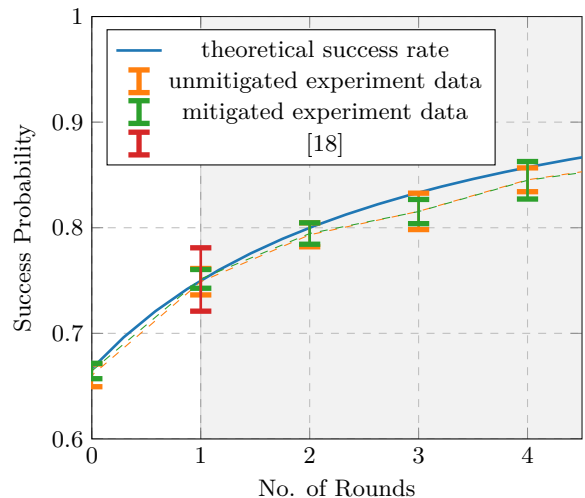
\begin{figure}[]
    \centering 

    \begin{tikzpicture}[scale=1]
        \begin{axis}[
            xlabel={No. of Rounds},
            ylabel={Success Probability},
            xmin=0, xmax=4.5,
            ymin=0.6, ymax=1,
            xtick={0,1, 2, 3, 4},
            domain = {0:7},
            ytick={0.6,0.7,0.8,0.9,1},
            legend pos=north west,
            ymajorgrids=true,
            xmajorgrids=true,
            grid style=dashed,
            error bars/y dir=both, 
            error bars/y explicit,  
            width=0.45\textwidth,
            height=0.4\textwidth
        ]   

        \filldraw[fill=lightgray,opacity=0.2] (100,0) rectangle (500,500);
        \addlegendimage{color1bg,line width = 1pt }
        \addlegendimage{error bar legend={color2bg}}
        \addlegendimage{error bar legend={color3bg}}
        \addlegendimage{error bar legend={color4bg}}

        \addplot[color1bg,line width = 1 pt] {(x+2)/(x+3)};
        
        \addplot[mark=nomark,color=color2bg, mark size=1pt,dashed, mark options={thin,solid},error bars/.cd, x dir=none, y dir=both, y explicit,
            error bar style={color=color2bg,solid, line width = 1 pt}, error mark options = {line width = 8 pt}] 
            table [x expr = \thisrow{no. of rounds}, y expr =\thisrow{distillation rate}, y error expr = \thisrow{std}, col sep=comma, mark=square] {./epr_dist_rate_v_rounds_aachen_nobarrier.csv};

        \addplot[mark=nomark,color=color3bg, mark size=1pt,dashed, mark options={thin,solid},error bars/.cd, x dir=none, y dir=both, y explicit,
            error bar style={color=color3bg,solid, line width = 1 pt}, error mark options = {line width = 8 pt}] 
            table [x expr = \thisrow{no. of rounds}, y expr =\thisrow{mitigated distillation rate}, y error expr = \thisrow{std}, col sep=comma, mark=square] {./epr_dist_rate_v_rounds_mit_aachen_nobarrier.csv};

        \addplot[mark=nomark,color=color4bg, mark size=1pt,dashed, mark options={thin,solid},error bars/.cd, x dir=none, y dir=both, y explicit,
            error bar style={color=color4bg,solid, line width = 1 pt}, error mark options = {line width = 8 pt}] 
            table[x=x,y=y,y error = error]{
            x y error 
            1 0.751 0.03
            };

        \addlegendentry{theoretical success rate}

        \addlegendentry{unmitigated experiment data}
        \addlegendentry{mitigated experiment data}

        \addlegendentry{\cite{OpticalSagnacRPD}}

        \end{axis}
    \end{tikzpicture}
    \hspace{15pt}
    \caption{The probability of successful distillation on \texttt{ibm\_aachen}. The blue line corresponds to the theoretical prediction, originally derived in \cite{OpticalSagnacRPD}. The orange and green datapoints correspond to unmitigated and mitigated with M3 respectively. Error bars corresponding to 1-$\sigma$ uncertainty were obtained by repeating experiments over five consecutive trials. Data outside the shaded grey region were obtained in a \textit{specific} party distillation protocol as opposed to a \textit{random} distillation protocol. We find a record success probability of 85\%, superior to both specific party distllation ($66\%$) and prior work ($75\%$) \cite{OpticalSagnacRPD}.}
    \label{fig:epr_dist_rate}
\end{figure}

We estimate the EPR distillation rate for up to four rounds of execution by counting the events in which one (and only one) of Alice, Bob, and Charlie measure an outcome of ``1" during a round of weak measurements to calculate the success probability given in \eqref{eq:bayes_success_probability}. Fig. \ref{fig:epr_dist_rate} shows the probability of success $P_N$, in which an entangled pair is distilled, and how it experimentally scales with increasing number of rounds. This result is compared to the theoretical prediction in \eqref{eq:success_prob_theory}.
Readout errors are mitigated using a matrix-free measurement error mitigation (M3) approach, as outlined in \cite{PRXQuantum.2.040326}. Within the first two rounds, the experimental estimation of the protocol's success rate tracks the theoretical prediction, \eqref{eq:success_prob_theory}, within 1-$\sigma$ error. 
Importantly, we find that the success probability of our protocol achieves a record rate of $85\%$ for a single copy of a \textit{W} state, when executing with four rounds of operation. This contrasts with the result of \cite{OpticalSagnacRPD}, which was limited to a 75\% success probability with a chance of false-positive distillation events (indicated by error bars that exceed theoretical limits in distillation rate).
The deviation between experiment and theory after the third round can be attributed to a degradation in the fidelity of the \textit{W} state at the beginning of each round. 
In Fig. 2, the estimated success rate of our protocol differs only slightly between the mitigated and unmitigated data. In the Supplementary Material, we show that measurement error mitigation becomes more important when using hardware with higher readout errors or when measuring outside the Z basis, as discussed later.

To confirm the degradation in \textit{W} state fidelity, we estimate the projector $\ket{W}\!\!\bra{W}$ after each stage of weak measurements. As shown in \eqref{eq:state_after_coupling}, the state after coupling to Alice, Bob, and Charlie's ancillary qubits contains a superposition of both the  $\ket{W}$ and $\ket{\Psi^+}$ states. Consequently, one must post-select their measurement results so that instances in which the three parties do not all measure ``0" are discarded. Estimation of an observable post-selected on mid-circuit measurement data is a commonplace in quantum computing experiments; all code used in these experiments is provided in \cite{conditional_measurement_toolbox}. 

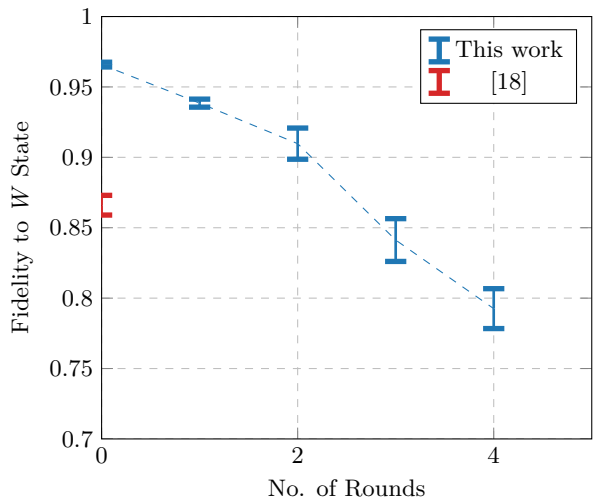
\begin{figure}[]
    \centering 

    \begin{tikzpicture}[scale=1]
        \begin{axis}[
            xlabel={No. of Rounds},
            ylabel={Fidelity to \textit{W} State},
            xmin=0, xmax=5,
            ymin=0.7, ymax=1,
            xtick={0,2,4},
            domain = {0:5},
            ytick={0.7,0.75,0.8,0.85,0.9,0.95,1},
            legend pos=north east,
            ymajorgrids=true,
            xmajorgrids=true,
            grid style=dashed,
            error bars/y dir=both, 
            error bars/y explicit,  
            width=0.45\textwidth,
            height=0.4\textwidth
        ]   
        \addlegendimage{error bar legend={color1bg}}
        \addlegendimage{error bar legend={color4bg}}
        \addplot[mark=nomark,color=color1bg, mark size=1pt,dashed, mark options={thin,solid},error bars/.cd, x dir=none, y dir=both, y explicit,
            error bar style={color=color1bg,solid, line width = 1 pt}, error mark options = {line width = 8 pt}] 
            table [x expr = \thisrow{no. of rounds}, y expr =\thisrow{w_exp_list_mit}, y error expr = \thisrow{w_std_mit}, col sep=comma, mark=square] {./w_fid_dd_notwirling_opt_qubits_opt2_nobarrier.csv};


        \addplot[mark=nomark,color=color2bg, mark size=1pt,dashed, mark options={thin,solid},error bars/.cd, x dir=none, y dir=both, y explicit,
            error bar style={color=color4bg,solid, line width = 1 pt}, error mark options = {line width = 8 pt}] 
            table[x=x,y=y,y error = error]{
            x y error 
            0 0.866 0.007
            };

        \addlegendentry{This work}

        \addlegendentry{\cite{OpticalSagnacRPD}}


        \end{axis}
    \end{tikzpicture}
    \hspace{15pt}
    \caption{The evolution of \textit{W} state fidelity (blue), determined by a series of measurements in the Pauli basis. Error bars corresponding to 1-$\sigma$ uncertainty were obtained by repeating experiments over five consecutive trials. The initial \textit{W} state fidelity of \cite{OpticalSagnacRPD} is shown in red for comparison. We find our W state fidelity to exceed that of \cite{OpticalSagnacRPD} even after two rounds of protocol execution.}
    \label{fig:w_state_fidelity_evolution}
\end{figure}

Fig. \ref{fig:w_state_fidelity_evolution} shows how the fidelity of Alice, Bob, and Charlie's \textit{W} state changes at the beginning of each round. We find the most dominant error mechanism to be the natural dephasing of qubits on IBM hardware \cite{PhysRevApplied.22.054074}. Dephasing is to be expected in circuits with many stages of mid-circuit measurements, since these often contain idle periods that are relatively long (836 ns) when compared to gate operations (20 - 60 ns). For a deeper discussion on the error mechanisms resulting from mid-circuit measurements in dynamic circuits, the reader is encouraged to consult \cite{PhysRevA.111.012611}. We employ XY4 dynamical decoupling \cite{ezzell2023dynamical} on idle qubits during periods of mid-circuit measurements. The XY4 sequence is by no means optimal, since it fails to account for cross-talk between idle qubits; further improvements may require ``staggered dynamical decoupling" between neighbouring qubits \cite{niu2024multi}. Despite the unavoidable degradation in the fidelity of the \textit{W} state, we find that the system shows better performance than \cite{OpticalSagnacRPD}, where the initial fidelity values were limited to 0.866, compared to the value of 0.96 obtained in this work.

Finally, to quantify the performance of our distillation protocol, we estimate a lower bound for the average entanglement of formation of EPR pairs extracted from each stage of protocol execution.
To estimate this lower bound, we use a method described by Bennett et al. in \cite{PhysRevA.54.3824} that only requires knowledge of the diagonal components of a density matrix in the Bell-basis. We shall refer to this representation as $\rho_{BDS}$, named after the fact that the state is a ``Bell-diagonal state":

\begin{equation}
    \rho_{BDS} = \sum_i\theta_i\ket{\Psi_i}\bra{\Psi_i},
\end{equation}

where $\ket{\Psi_i}$ refer to the four Bell-states (i.e. $\ket{\Psi_i}\in \{\ket{\Psi^+},\ket{\Psi^-},\ket{\Phi^+},\ket{\Phi^-}\}$) and $\theta_i$ are real-valued weights that sum to one. Intuitively, the weights $\theta_i$ also correspond to state fidelities; i.e. $\theta_i ~=~ \bra{\Psi_i}\rho_{BDS}\ket{\Psi_i}$.

Bennett et al. introduce the largest ``fully-entangled fraction" $\theta_{max} = \max_i\theta_i$ as the largest fidelity between a given state and any Bell-state. One can define the function $f(\theta_{max})$:

\begin{equation}
    f(\theta_{max}) = 
    \begin{dcases}
       h(1/2 - \sqrt{\theta_{max}(1-\theta_{max})}) & \theta_{max} > 1/2 \\
       0 & \theta_{max} \leq 1/2
    \end{dcases},
    \label{eq:eof_lowerbound}
\end{equation}

which is proven in \cite{PhysRevA.54.3824} to be a lower bound for the entanglement of formation $E(\rho)$ for \textit{any} spin-1/2 system $\rho$. The bound reaches equality only when $\rho$ is a Bell-diagonal state. This result implies that with only knowledge of the fidelity to the closest Bell state, one can calculate a lower bound to $E(\rho)$ when $\theta_{max}>1/2$. 

\begin{figure}[]
    \centering 

    \begin{tikzpicture}[scale=1]
        \begin{axis}[
            xlabel={No. of Rounds},
            ylabel={Fidelity to $\Psi^+$ State},
            xmin=0, xmax=5,
            ymin=0.5, ymax=1,
            xtick={0,1,2,3,4,5},
            domain = {0:7},
            ytick={0.5,0.6,0.7,0.8,0.9,1},
            legend pos=north east,
            ymajorgrids=true,
            xmajorgrids=true,
            grid style=dashed,
            axis y line* = left,
            y axis line style={color1bg},
            ylabel style ={color1bg},
            every y tick label/.append style={color1bg},
            error bars/y dir=both, 
            error bars/y explicit,  
            width=0.45\textwidth,
            height=0.4\textwidth
        ]   
        \node at (axis cs:0.5,0.75) [anchor=north east,rotate=90] {Specific Party};
        \node at (axis cs:1,0.75) [anchor=north east,rotate=90] {Random Party};
        \filldraw[fill=lightgray,opacity=0.2] (100,0) rectangle (500,500);
        
        \addplot[mark=nomark,color=color1bg, mark size=1pt,dashed, mark options={thin,solid},error bars/.cd, x dir=none, y dir=both, y explicit,
            error bar style={color=color1bg,solid, line width = 1 pt}, error mark options = {line width = 8 pt}] 
            table [x expr = \thisrow{no. of rounds}, y expr =\thisrow{avg_fidelity_mit}, y error expr = \thisrow{avg_fidelity_mit_std}, col sep=comma, mark=square] {./avg_epr_fidelity_with_eof_mc.csv}; \label{fidelity}

        \end{axis}
        \pgfplotsset{every axis y label/.append style={rotate=180}}
          \begin{axis}[
                xmin=0, xmax=5,
                ymin = 0, ymax=1,
                ytick={0,0.2,0.4,0.6,0.8,1},
                hide x axis,
                axis y line*=right,
                ylabel={$\frac{1}{|\mathcal{S}|}\sum_iE(\rho_i)$},
                y axis line style={color2bg},
                ylabel style ={color2bg},
                every y tick label/.append style={color2bg},
                width=0.45\textwidth,
            height=0.4\textwidth
            ]
            \addlegendimage{error bar legend={color1bg}}
            \addlegendimage{error bar legend={color2bg}}
            \addlegendimage{/pgfplots/refstyle=fidelity}\addlegendentry{Fidelity Estimation}

            \addplot[mark=nomark,color=color2bg, mark size=1pt,dashed, mark options={thin,solid},error bars/.cd, x dir=none, y dir=both, y explicit,
            error bar style={color=color2bg,solid, line width = 1 pt}, error mark options = {line width = 8 pt}] 
            table [x expr = \thisrow{no. of rounds}, y expr =\thisrow{eof_lower_bound_mc_mean_mit}, y error expr = \thisrow{eof_lower_bound_mc_std_mit}, col sep=comma, mark=square] {./avg_epr_fidelity_with_eof_mc.csv};

            \addlegendentry{Entanglement of Formation}

            
          \end{axis}

    \end{tikzpicture}
    \hspace{15pt}
    \caption{The evolution of EPR state fidelity and average entanglement of formation (lower bound), measured by a projection onto the $\Psi^+$ state in the Pauli basis. Error bars corresponding to 1-$\sigma$ uncertainty were obtained by repeating experiments over five consecutive trials. Data outside the shaded grey region were obtained in a \textit{specific} party distillation protocol as opposed to a \textit{random} distillation protocol. We observe a deterioration in state fidelity, placing limits on the achievable expected entanglement $\expval{E}$.}
    \label{fig:epr_fid_eof}
\end{figure}
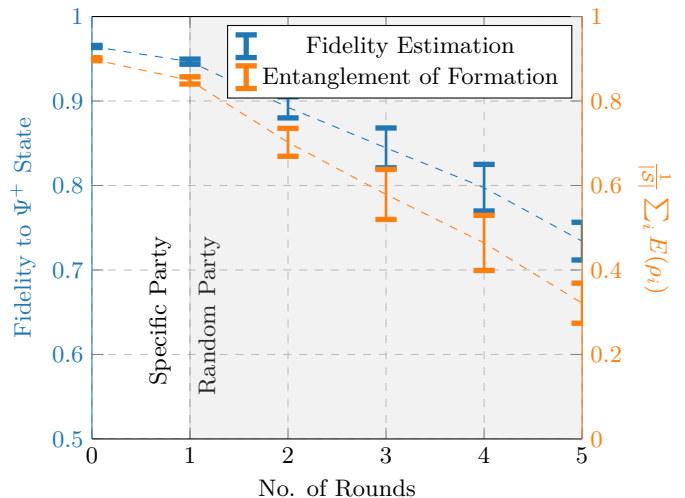

Fig. \ref{fig:epr_fid_eof} shows the evolution of both distilled entanglement of formation, and fidelity to the triplet state. All data presented were recorded using an XY4 dynamical decoupling sequence during idle periods. We attribute the non-ideal values of entanglement of formation to the imperfect \textit{W} states used at the beginning of each round and finite gate errors.  Weighting the success probabilities shown in Fig. \ref{fig:epr_dist_rate} according to the corresponding entanglement of formation values, we find the average distilled entanglement of our protocol to be as high as $\expval{E} \geq   0.6379$ pairs/\textit{W} state, when using a single-round implementation of the protocol and as low as $\expval{E}\geq 0.5404$ for a four-round implementation. The deterioration in the lower bounds of $\expval{E}$ is attributed to a corresponding degradation in the fidelity of the distilled pair (Fig. \ref{fig:epr_fid_eof}); this is especially damaging to performance, since the probability of success per round decreases as the total number of rounds increases. On the same hardware, we find that $\expval{E} \geq 0.5993$ is obtained by using specific party distillation. We have thus shown the first experimental demonstration of an advantage in distilled entanglement rate \textit{per copy} of random party versus specific party distillation. Further improvements in the performance of the random party distillation rate depend on decoupling from external dephasing channels and having shorter readout periods to minimize errors introduced during mid-circuit measurements. 

\section{Conclusion}

In this work, we have presented the first physical demonstration of random party distillation that exceeds one round of execution, thus achieving a record distillation rate as a result. We have also shown a scenario in which a random party distillation protocol using weak measurements outperforms its counterpart using strong measurements alone. As a consequence, our work serves as a proof of concept for utilizing weak measurements in the regime of few copies for bipartite entanglement distribution (or localization, as in \cite{PhysRevA.71.042306}). Indeed, a natural continuation of this work may consider few-copy implementations of tasks such as gate teleportation or localization of computational resources, such as magic, or ``non-stabilizerness \cite{bravyi2005universal}."

Moreover, we have also shown how the dephasing of superconducting qubits impacts the quality of both bi- and multipartite entanglement which can be partially mitigated using a combination of dynamical decoupling and M3. More broadly, the results of this work point to the importance of short readout lengths for the execution of either dynamic circuits or those that employ mid-circuit measurements. Such a change would require an entirely new readout architecture, an example being the method of ``junction readout" \cite{chapple2025balancedcrosskerrcouplingsuperconducting}. In the meantime, it is important to explore alternative methods of dynamical decoupling with the goal of preserving multi-qubit entanglement in mind with techniques such as staggered dynamical decoupling \cite{niu2024multi} and using novel schemes obtained with statistical learning methods \cite{tong2025empiricallearningdynamicaldecoupling}. 

Alternatively, using a device with greater connectivity between qubits would allow the user to use a different set of ancillary qubits with each round of weak measurements. In doing so, one could defer mid-circuit measurements at the cost of greater overhead in terms of the number of qubits required for bipartite distillation. A presentation of this alternative protocol can be found in the supplementary material to this work.

As quantum computing moves toward large-scale distributed architectures \cite{mohseni2025buildquantumsupercomputerscaling}, the future of random party
distillation extends beyond immediate improvements in
distillation rates and circuit techniques. Few-copy protocols, such as the 
random party distillation we have shown in this work, could play an important role in generating high-fidelity entanglement between nodes in a quantum network, a necessary step for enabling robust and
fault-tolerant distributed computation \cite{carrera2024combining}.  


\section{Acknowledgements}
The authors thank Dr. Brian T. Kirby and Dr. Sean Wagner for insightful discussions. The authors thank PINQ2 for providing access to the superconducting processors used in this work. 

The authors acknowledge the following funding sources: Horizon Europe - Canada Project 101070168 HYPERSPACE, NSERC ALLRP 578462 - 22 and 578460 - 22, CFI and ORF Project 33415, NSERC Discovery Funds RGPIN-2019-07019, RGPAS-2019-00113, and RGPIN-2025-05068. Finally, the authors acknowledge support from the NUS start-up and CQT PI grants. 


\bibliography{random_party.bib}

@article{QuantEntangle,
  title = {Quantum entanglement},
  author = {Horodecki, Ryszard and Horodecki, Pawe\l{} and Horodecki, Micha\l{} and Horodecki, Karol},
  journal = {Rev. Mod. Phys.},
  volume = {81},
  issue = {2},
  pages = {865--942},
  numpages = {0},
  year = {2009},
  month = {Jun},
  publisher = {American Physical Society},
  doi = {10.1103/RevModPhys.81.865},
  url = {https://link.aps.org/doi/10.1103/RevModPhys.81.865}
}

@article{OpticalSagnacRPD,
  title = {Experimental random-party entanglement distillation via weak measurement},
  author = {Li, Zheng-Da and Yuan, Xiao and Yin, Xu-Fei and Liu, Li-Zheng and Zhang, Rui and Fei, Yue-Yang and Li, Li and Liu, Nai-Le and Ma, Xiongfeng and Lu, He and Chen, Yu-Ao and Pan, Jian-Wei},
  journal = {Phys. Rev. Res.},
  volume = {2},
  issue = {2},
  pages = {023047},
  numpages = {9},
  year = {2020},
  month = {Apr},
  publisher = {American Physical Society},
  doi = {10.1103/PhysRevResearch.2.023047},
  url = {https://link.aps.org/doi/10.1103/PhysRevResearch.2.023047}
}

@article{QutritRPDTheory,
  title = {Random Bipartite Entanglement from $W$ and $W$-Like States},
  author = {Fortescue, Ben and Lo, Hoi-Kwong},
  journal = {Phys. Rev. Lett.},
  volume = {98},
  issue = {26},
  pages = {260501},
  numpages = {4},
  year = {2007},
  month = {Jun},
  publisher = {American Physical Society},
  doi = {10.1103/PhysRevLett.98.260501},
  url = {https://link.aps.org/doi/10.1103/PhysRevLett.98.260501}
}

@article{Liu2023roundcomplexity,
  doi = {10.22331/q-2023-09-07-1104},
  url = {https://doi.org/10.22331/q-2023-09-07-1104},
  title = {The {R}ound {C}omplexity of {L}ocal {O}perations and {C}lassical {C}ommunication ({LOCC}) in {R}andom-{P}arty {E}ntanglement {D}istillation},
  author = {Liu, Guangkuo and George, Ian and Chitambar, Eric},
  journal = {{Quantum}},
  issn = {2521-327X},
  publisher = {{Verein zur F{\"{o}}rderung des Open Access Publizierens in den Quantenwissenschaften}},
  volume = {7},
  pages = {1104},
  month = sep,
  year = {2023}
}

@article{PhysRevLett.98.260501,
  title = {Random Bipartite Entanglement from $W$ and $W$-Like States},
  author = {Fortescue, Ben and Lo, Hoi-Kwong},
  journal = {Phys. Rev. Lett.},
  volume = {98},
  issue = {26},
  pages = {260501},
  numpages = {4},
  year = {2007},
  month = {Jun},
  publisher = {American Physical Society},
  doi = {10.1103/PhysRevLett.98.260501},
  url = {https://link.aps.org/doi/10.1103/PhysRevLett.98.260501}
}

@article{wang2024randomdistillationprotocolslong,
      title={Random distillation protocols in long baseline telescopy},
      author={Wang, Yunkai and Chitambar, Eric},
      journal={Physical review letters},
      volume={134},
      number={17},
      pages={170801},
      year={2025},
      publisher={APS}
}

@article{PRXQuantum.2.040326,
  title = {Scalable Mitigation of Measurement Errors on Quantum Computers},
  author = {Nation, Paul D. and Kang, Hwajung and Sundaresan, Neereja and Gambetta, Jay M.},
  journal = {PRX Quantum},
  volume = {2},
  issue = {4},
  pages = {040326},
  numpages = {9},
  year = {2021},
  month = {Nov},
  publisher = {American Physical Society},
  doi = {10.1103/PRXQuantum.2.040326},
  url = {https://link.aps.org/doi/10.1103/PRXQuantum.2.040326}
}

@misc{mohseni2025buildquantumsupercomputerscaling,
      title={How to Build a Quantum Supercomputer: Scaling from Hundreds to Millions of Qubits}, 
      author={Masoud Mohseni and Artur Scherer and K. Grace Johnson and Oded Wertheim and Matthew Otten and Navid Anjum Aadit and Yuri Alexeev and Kirk M. Bresniker and Kerem Y. Camsari and Barbara Chapman and Soumitra Chatterjee and Gebremedhin A. Dagnew and Aniello Esposito and Farah Fahim and Marco Fiorentino and Archit Gajjar and Abdullah Khalid and Xiangzhou Kong and Bohdan Kulchytskyy and Elica Kyoseva and Ruoyu Li and P. Aaron Lott and Igor L. Markov and Robert F. McDermott and Giacomo Pedretti and Pooja Rao and Eleanor Rieffel and Allyson Silva and John Sorebo and Panagiotis Spentzouris and Ziv Steiner and Boyan Torosov and Davide Venturelli and Robert J. Visser and Zak Webb and Xin Zhan and Yonatan Cohen and Pooya Ronagh and Alan Ho and Raymond G. Beausoleil and John M. Martinis},
      year={2025},
      eprint={2411.10406},
      archivePrefix={arXiv},
      primaryClass={quant-ph},
      url={https://arxiv.org/abs/2411.10406}, 
}

@article{PhysRevA.78.012348,
  title = {Random-party entanglement distillation in multiparty states},
  author = {Fortescue, Ben and Lo, Hoi-Kwong},
  journal = {Phys. Rev. A},
  volume = {78},
  issue = {1},
  pages = {012348},
  numpages = {9},
  year = {2008},
  month = {Jul},
  publisher = {American Physical Society},
  doi = {10.1103/PhysRevA.78.012348},
  url = {https://link.aps.org/doi/10.1103/PhysRevA.78.012348}
}

@article{PRXQuantum.5.030339,
  title = {Efficient Long-Range Entanglement Using Dynamic Circuits},
  author = {B\"aumer, Elisa and Tripathi, Vinay and Wang, Derek S. and Rall, Patrick and Chen, Edward H. and Majumder, Swarnadeep and Seif, Alireza and Minev, Zlatko K.},
  journal = {PRX Quantum},
  volume = {5},
  issue = {3},
  pages = {030339},
  numpages = {20},
  year = {2024},
  month = {Aug},
  publisher = {American Physical Society},
  doi = {10.1103/PRXQuantum.5.030339},
  url = {https://link.aps.org/doi/10.1103/PRXQuantum.5.030339}
}

@article{PhysRevApplied.23.014057,
  title = {Teleporting two-qubit entanglement across 19 qubits on a superconducting quantum computer},
  author = {Kang, Haiyue and Kam, John F. and Mooney, Gary J. and Hollenberg, Lloyd C.L.},
  journal = {Phys. Rev. Appl.},
  volume = {23},
  issue = {1},
  pages = {014057},
  numpages = {17},
  year = {2025},
  month = {Jan},
  publisher = {American Physical Society},
  doi = {10.1103/PhysRevApplied.23.014057},
  url = {https://link.aps.org/doi/10.1103/PhysRevApplied.23.014057}}

@article{PhysRevA.103.042605,
  title = {Mitigating measurement errors in multiqubit experiments},
  author = {Bravyi, Sergey and Sheldon, Sarah and Kandala, Abhinav and Mckay, David C. and Gambetta, Jay M.},
  journal = {Phys. Rev. A},
  volume = {103},
  issue = {4},
  pages = {042605},
  numpages = {12},
  year = {2021},
  month = {Apr},
  publisher = {American Physical Society},
  doi = {10.1103/PhysRevA.103.042605},
  url = {https://link.aps.org/doi/10.1103/PhysRevA.103.042605}
}

@article{niu2024multi,
  title={Multi-qubit dynamical decoupling for enhanced crosstalk suppression},
  author={Niu, Siyuan and Todri-Sanial, Aida and Bronn, Nicholas T},
  journal={Quantum Science and Technology},
  volume={9},
  number={4},
  pages={045003},
  year={2024},
  publisher={IOP Publishing}
}

@misc{tong2025empiricallearningdynamicaldecoupling,
      title={Empirical learning of dynamical decoupling on quantum processors}, 
      author={Christopher Tong and Helena Zhang and Bibek Pokharel},
      year={2025},
      eprint={2403.02294},
      archivePrefix={arXiv},
      primaryClass={quant-ph},
      url={https://arxiv.org/abs/2403.02294}, 
}

@article{eisert2007quantitative,
  title={Quantitative entanglement witnesses},
  author={Eisert, Jens and Brandao, Fernando GSL and Audenaert, Koenraad MR},
  journal={New Journal of Physics},
  volume={9},
  number={3},
  pages={46},
  year={2007},
  publisher={IOP Publishing}
}

@article{PhysRevLett.80.2245,
  title = {Entanglement of Formation of an Arbitrary State of Two Qubits},
  author = {Wootters, William K.},
  journal = {Phys. Rev. Lett.},
  volume = {80},
  issue = {10},
  pages = {2245--2248},
  numpages = {0},
  year = {1998},
  month = {Mar},
  publisher = {American Physical Society},
  doi = {10.1103/PhysRevLett.80.2245},
  url = {https://link.aps.org/doi/10.1103/PhysRevLett.80.2245}
}

@article{PhysRevA.72.042318,
  title = {Mixed-state entanglement of assistance and the generalized concurrence},
  author = {Gour, Gilad},
  journal = {Phys. Rev. A},
  volume = {72},
  issue = {4},
  pages = {042318},
  numpages = {7},
  year = {2005},
  month = {Oct},
  publisher = {American Physical Society},
  doi = {10.1103/PhysRevA.72.042318},
  url = {https://link.aps.org/doi/10.1103/PhysRevA.72.042318}
}

@misc{chapple2025balancedcrosskerrcouplingsuperconducting,
      title={Balanced cross-Kerr coupling for superconducting qubit readout}, 
      author={Alex A. Chapple and Othmane Benhayoune-Khadraoui and Simon Richer and Alexandre Blais},
      year={2025},
      eprint={2501.09010},
      archivePrefix={arXiv},
      primaryClass={quant-ph},
      url={https://arxiv.org/abs/2501.09010}, 
}

@software{conditional_measurement_toolbox,author = {Greenwood, Alexander and Russett, Jackson},month = mar,title = {{Conditional Measurement Toolbox}},url = {https://github.com/alexg-jpg/conditional_measurements},version = {1.0.0},year = {2025}}

@article{PhysRevA.85.062316,
  title = {Entanglement monotones for $W$-type states},
  author = {Chitambar, Eric and Cui, Wei and Lo, Hoi-Kwong},
  journal = {Phys. Rev. A},
  volume = {85},
  issue = {6},
  pages = {062316},
  numpages = {10},
  year = {2012},
  month = {Jun},
  publisher = {American Physical Society},
  doi = {10.1103/PhysRevA.85.062316},
  url = {https://link.aps.org/doi/10.1103/PhysRevA.85.062316}
}

@inproceedings{enriquez2016maximally,
  title={Maximally entangled multipartite states: a brief survey},
  author={Enr{\'\i}quez, M and Wintrowicz, I and {\.Z}yczkowski, Karol},
  booktitle={Journal of Physics: Conference Series},
  volume={698},
  number={1},
  pages={012003},
  year={2016},
  organization={IOP Publishing}
}

@article{PhysRevApplied.22.054074,
  title = {Learning how to dynamically decouple by optimizing rotational gates},
  author = {Rahman, Arefur and Egger, Daniel J. and Arenz, Christian},
  journal = {Phys. Rev. Appl.},
  volume = {22},
  issue = {5},
  pages = {054074},
  numpages = {12},
  year = {2024},
  month = {Nov},
  publisher = {American Physical Society},
  doi = {10.1103/PhysRevApplied.22.054074},
  url = {https://link.aps.org/doi/10.1103/PhysRevApplied.22.054074}
}

@article{Kanazawa_Qiskit_Experiments_A_2023,
author = {Kanazawa, Naoki and Egger, Daniel J. and Ben-Haim, Yael and Zhang, Helena and Shanks, William E. and Aleksandrowicz, Gadi and Wood, Christopher J.},
doi = {10.21105/joss.05329},
journal = {Journal of Open Source Software},
month = apr,
number = {84},
pages = {5329},
title = {{Qiskit Experiments: A Python package to characterize and calibrate quantum computers}},
url = {https://joss.theoj.org/papers/10.21105/joss.05329},
volume = {8},
year = {2023}
}

@article{PhysRevLett.127.140501,
  title = {Computable and Operationally Meaningful Multipartite Entanglement Measures},
  author = {Beckey, Jacob L. and Gigena, N. and Coles, Patrick J. and Cerezo, M.},
  journal = {Phys. Rev. Lett.},
  volume = {127},
  issue = {14},
  pages = {140501},
  numpages = {7},
  year = {2021},
  month = {Sep},
  publisher = {American Physical Society},
  doi = {10.1103/PhysRevLett.127.140501},
  url = {https://link.aps.org/doi/10.1103/PhysRevLett.127.140501}
}

@article{PhysRevA.111.032423,
  title = {Entanglement of assistance as a measure of multiparty entanglement},
  author = {Biswas, Indranil and Bhunia, Atanu and Bera, Subrata and Chattopadhyay, Indrani and Sarkar, Debasis},
  journal = {Phys. Rev. A},
  volume = {111},
  issue = {3},
  pages = {032423},
  numpages = {10},
  year = {2025},
  month = {Mar},
  publisher = {American Physical Society},
  doi = {10.1103/PhysRevA.111.032423},
  url = {https://link.aps.org/doi/10.1103/PhysRevA.111.032423}
}

@article{carrera2024combining,
  title={Combining quantum processors with real-time classical communication},
  author={Carrera Vazquez, Almudena and Tornow, Caroline and Rist{\`e}, Diego and Woerner, Stefan and Takita, Maika and Egger, Daniel J},
  journal={Nature},
  pages={1--5},
  year={2024},
  publisher={Nature Publishing Group UK London}
}

@article{PhysRevLett.108.240504,
  title = {Increasing Entanglement Monotones by Separable Operations},
  author = {Chitambar, Eric and Cui, Wei and Lo, Hoi-Kwong},
  journal = {Phys. Rev. Lett.},
  volume = {108},
  issue = {24},
  pages = {240504},
  numpages = {4},
  year = {2012},
  month = {Jun},
  publisher = {American Physical Society},
  doi = {10.1103/PhysRevLett.108.240504},
  url = {https://link.aps.org/doi/10.1103/PhysRevLett.108.240504}
}

@article{PhysRevA.54.3824,
  title = {Mixed-state entanglement and quantum error correction},
  author = {Bennett, Charles H. and DiVincenzo, David P. and Smolin, John A. and Wootters, William K.},
  journal = {Phys. Rev. A},
  volume = {54},
  issue = {5},
  pages = {3824--3851},
  numpages = {0},
  year = {1996},
  month = {Nov},
  publisher = {American Physical Society},
  doi = {10.1103/PhysRevA.54.3824},
  url = {https://link.aps.org/doi/10.1103/PhysRevA.54.3824}
}

@article{sun2017entanglement,
  title={Entanglement swapping over 100 km optical fiber with independent entangled photon-pair sources},
  author={Sun, Qi-Chao and Jiang, Yang-Fan and Mao, Ya-Li and You, Li-Xing and Zhang, Wei and Zhang, Wei-Jun and Jiang, Xiao and Chen, Teng-Yun and Li, Hao and Huang, Yi-Dong and others},
  journal={Optica},
  volume={4},
  number={10},
  pages={1214--1218},
  year={2017},
  publisher={Optical Society of America}
}

@article{PhysRevA.84.052301,
  title = {Randomly distilling $W$-class states into general configurations of two-party entanglement},
  author = {Cui, Wei and Chitambar, Eric and Lo, Hoi-Kwong},
  journal = {Phys. Rev. A},
  volume = {84},
  issue = {5},
  pages = {052301},
  numpages = {11},
  year = {2011},
  month = {Nov},
  publisher = {American Physical Society},
  doi = {10.1103/PhysRevA.84.052301},
  url = {https://link.aps.org/doi/10.1103/PhysRevA.84.052301}
}

@article{PhysRevLett.126.190503,
  title = {Experimentally Accessible Bounds on Distillable Entanglement from Entropic Uncertainty Relations},
  author = {Bergh, Bjarne and G\"arttner, Martin},
  journal = {Phys. Rev. Lett.},
  volume = {126},
  issue = {19},
  pages = {190503},
  numpages = {6},
  year = {2021},
  month = {May},
  publisher = {American Physical Society},
  doi = {10.1103/PhysRevLett.126.190503},
  url = {https://link.aps.org/doi/10.1103/PhysRevLett.126.190503}
}

@article{lo2001proof,
  title={Proof of unconditional security of six-state quatum key distribution scheme},
  author={Lo, Hoi-Kwong},
  journal={Quantum Information \& Computation},
  volume={1},
  number={2},
  pages={81--94},
  year={2001},
  publisher={Rinton Press, Incorporated Paramus, NJ}
}

@article{PhysRevA.111.012611,
  title = {Randomized benchmarking protocol for dynamic circuits},
  author = {Shirizly, Liran and Govia, Luke C. G. and McKay, David C.},
  journal = {Phys. Rev. A},
  volume = {111},
  issue = {1},
  pages = {012611},
  numpages = {12},
  year = {2025},
  month = {Jan},
  publisher = {American Physical Society},
  doi = {10.1103/PhysRevA.111.012611},
  url = {https://link.aps.org/doi/10.1103/PhysRevA.111.012611}
}

@article{ezzell2023dynamical,
  title={Dynamical decoupling for superconducting qubits: A performance survey},
  author={Ezzell, Nic and Pokharel, Bibek and Tewala, Lina and Quiroz, Gregory and Lidar, Daniel A},
  journal={Physical Review Applied},
  volume={20},
  number={6},
  pages={064027},
  year={2023},
  publisher={APS}
}

@article{roy2025parity,
  title={Parity-dependent state transfer for direct entanglement generation},
  author={Roy, Federico A and Romeiro, Jo{\~a}o H and Koch, Leon and Tsitsilin, Ivan and Schirk, Johannes and Glaser, Niklas J and Bruckmoser, Niklas and Singh, Malay and Haslbeck, Franz X and Huber, Gerhard BP and others},
  journal={Nature Communications},
  volume={16},
  number={1},
  pages={2660},
  year={2025},
  publisher={Nature Publishing Group UK London}
}

@article{PhysRevA.71.042306,
  title = {Localizable entanglement},
  author = {Popp, M. and Verstraete, F. and Mart\'{\i}n-Delgado, M. A. and Cirac, J. I.},
  journal = {Phys. Rev. A},
  volume = {71},
  issue = {4},
  pages = {042306},
  numpages = {18},
  year = {2005},
  month = {Apr},
  publisher = {American Physical Society},
  doi = {10.1103/PhysRevA.71.042306},
  url = {https://link.aps.org/doi/10.1103/PhysRevA.71.042306}
}

@article{PRXQuantum.5.010202,
  title = {Remote-Entanglement Protocols for Stationary Qubits with Photonic Interfaces},
  author = {Beukers, Hans K.C. and Pasini, Matteo and Choi, Hyeongrak and Englund, Dirk and Hanson, Ronald and Borregaard, Johannes},
  journal = {PRX Quantum},
  volume = {5},
  issue = {1},
  pages = {010202},
  numpages = {27},
  year = {2024},
  month = {Mar},
  publisher = {American Physical Society},
  doi = {10.1103/PRXQuantum.5.010202},
  url = {https://link.aps.org/doi/10.1103/PRXQuantum.5.010202}
}

@article{PhysRevLett.105.230502,
  title = {Iterative Entanglement Distillation: Approaching the Elimination of Decoherence},
  author = {Hage, Boris and Samblowski, Aiko and DiGuglielmo, James and Fiur\'a\ifmmode \check{s}\else \v{s}\fi{}ek, Jarom\'{\i}r and Schnabel, Roman},
  journal = {Phys. Rev. Lett.},
  volume = {105},
  issue = {23},
  pages = {230502},
  numpages = {4},
  year = {2010},
  month = {Dec},
  publisher = {American Physical Society},
  doi = {10.1103/PhysRevLett.105.230502},
  url = {https://link.aps.org/doi/10.1103/PhysRevLett.105.230502}
}

@article{abdelkhalek2016efficient,
  title={Efficient entanglement distillation without quantum memory},
  author={Abdelkhalek, Daniela and Syllwasschy, Mareike and Cerf, Nicolas J and Fiur{\'a}{\v{s}}ek, Jarom{\'\i}r and Schnabel, Roman},
  journal={Nature communications},
  volume={7},
  number={1},
  pages={11720},
  year={2016},
  publisher={Nature Publishing Group UK London}
}

@article{bravyi2005universal,
  title={Universal quantum computation with ideal Clifford gates and noisy ancillas},
  author={Bravyi, Sergey and Kitaev, Alexei},
  journal={Physical Review A—Atomic, Molecular, and Optical Physics},
  volume={71},
  number={2},
  pages={022316},
  year={2005},
  publisher={APS}
}

@article{PhysRevA.62.062314,
  title = {Three qubits can be entangled in two inequivalent ways},
  author = {D\"ur, W. and Vidal, G. and Cirac, J. I.},
  journal = {Phys. Rev. A},
  volume = {62},
  issue = {6},
  pages = {062314},
  numpages = {12},
  year = {2000},
  month = {Nov},
  publisher = {American Physical Society},
  doi = {10.1103/PhysRevA.62.062314},
  url = {https://link.aps.org/doi/10.1103/PhysRevA.62.062314}
}

@article{PhysRevLett.87.040401,
  title = {Classification of Mixed Three-Qubit States},
  author = {Ac\'{\i}n, A. and Bru\ss{}, D. and Lewenstein, M. and Sanpera, A.},
  journal = {Phys. Rev. Lett.},
  volume = {87},
  issue = {4},
  pages = {040401},
  numpages = {4},
  year = {2001},
  month = {Jul},
  publisher = {American Physical Society},
  doi = {10.1103/PhysRevLett.87.040401},
  url = {https://link.aps.org/doi/10.1103/PhysRevLett.87.040401}
}

@article{kretschmann2008information,
  title={The information-disturbance tradeoff and the continuity of Stinespring's representation},
  author={Kretschmann, Dennis and Schlingemann, Dirk and Werner, Reinhard F},
  journal={IEEE transactions on information theory},
  volume={54},
  number={4},
  pages={1708--1717},
  year={2008},
  publisher={IEEE}
}

\end{document}


\title{Supplementary Information: Random Party Distillation on a Superconducting Quantum Processor}

\maketitle



\section{System Benchmarking}

In this supplementary material, we provide device calibration data (Section I), additional protocol statistics (Section II), an alternative implementation of the protocol without mid-circuit measurements (Section III), and a simplified noise model used to interpret the experimental trends reported in the main text (Section IV). The experiments in the main text were performed on \texttt{ibm\_aachen}, while comparison data were also collected on \texttt{ibm\_quebec}. The corresponding qubit parameters at the time of data collection are listed in Tables~\ref{tab:qubits_aachen} and \ref{tab:qubits}.



Although \texttt{ibm\_quebec} exhibits longer average coherence times and shorter readout durations, \texttt{ibm\_aachen} provides lower average readout (latest readout error rates of $ 0.0046 <  0.0179$, see Tables \ref{tab:qubits_aachen} and \ref{tab:qubits}) and entangling-gate errors (latest error rates of $ 0.001 <  0.005$) for the qubits used in this work. As shown later in this section, \texttt{ibm\_aachen}'s characteristics are more favorable for multi-round random party distillation, where repeated entangling operations and measurements strongly influence the protocol performance.


\begin{table}[H]
    \centering
    \begin{tabular}{p{4em}p{9em}p{9em}p{9em}p{9em}p{9em}}
        \hline\hline
        \textbf{Qubit} & \textbf{Physical Qubit} & \textbf{Coherence\!\;\;Time $T_1$ ($\mu$s)} & \textbf{Dephasing\!\;\;Time $T_2$ ($\mu$s)} & \textbf{Readout Error}  & \textbf{Readout\!\;\;length (ns)} \\ \hline
         a & 24 & 222.4184 & 342.0892 & 0.0078 & 2584.0 \\
         b & 25 & 241.6378 & 230.3764 & 0.0071 & 2584.0 \\
         c & 26 & 253.8337 & 272.7626 & 0.0061 & 2584.0 \\
         $\alpha$ & 23 & 261.9766 & 296.9496 & 0.0033 & 2584.0 \\
         $\beta$ & 37 & 194.8966 & 184.6699 & 0.0061 & 2584.0 \\
         $\gamma$ & 27 & 264.3315 & 311.4086 & 0.0042 & 2584.0 \\
    \end{tabular}
    \caption{Physical parameters pertinent to the experiments presented in the manuscript. All parameters were taken from the superconducting device, \texttt{ibm\_aachen}. In this work, we attribute the improvements in random distillation rate to the superior readout and entangling gate errors exhibited by \texttt{ibm\_aachen} when compared against \texttt{ibm\_quebec}.}
    \label{tab:qubits_aachen}
\end{table}

\begin{table}[H]
    \centering
    \begin{tabular}{p{4em}p{9em}p{9em}p{9em}p{9em}p{9em}}
        \hline\hline
        \textbf{Qubit} & \textbf{Physical Qubit} & \textbf{Coherence\!\;\;Time $T_1$ ($\mathbf{\mu}$s)} & \textbf{Dephasing\!\;\;Time $T_2$ ($\mu$s)} & \textbf{Readout Error}  & \textbf{Readout\!\;\;length (ns)} \\ \hline
         a & 23 & 366.1082 & 216.9724  &  0.0100
 & 835.5556 \\
         b & 24 & 382.6895 & 182.1576  & 0.0283
 & 835.5556 \\
         c & 25 & 306.8350 & 482.8157 & 0.0066
 & 835.5556 \\
         $\alpha$ & 22 & 371.1372 & 288.1581 &  0.0244
 & 835.5556 \\
         $\beta$ & 34 & 282.4133 & 163.5965  & 0.0310
 & 835.5556 \\
         $\gamma$ & 26  & 399.3776 & 414.6890 & 0.0259
  & 835.5556\\
    \end{tabular}
    \caption{Physical parameters pertinent to the experiments presented in the manuscript. All parameters were taken from the superconducting device, \texttt{ibm\_quebec}.}
    \label{tab:qubits}
\end{table}

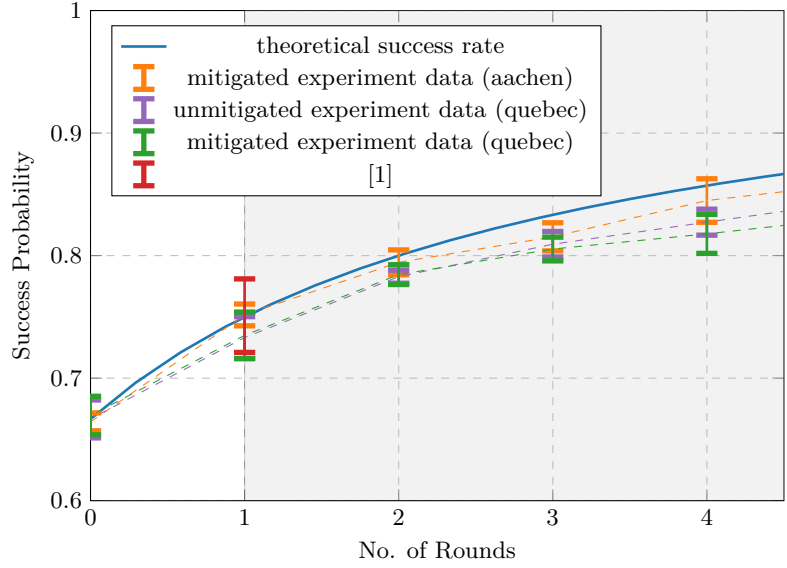
\begin{figure}[]
    \centering 

    \begin{tikzpicture}[scale=1]
        \begin{axis}[
            xlabel={No. of Rounds},
            ylabel={Success Probability},
            xmin=0, xmax=4.5,
            ymin=0.6, ymax=1,
            xtick={0,1, 2, 3, 4},
            domain = {0:7},
            ytick={0.6,0.7,0.8,0.9,1},
            legend pos=north west,
            ymajorgrids=true,
            xmajorgrids=true,
            grid style=dashed,
            error bars/y dir=both, 
            error bars/y explicit,  
            width=0.6\textwidth,
            height=0.45\textwidth
        ]   

        \filldraw[fill=lightgray,opacity=0.2] (100,0) rectangle (500,500);
        \addlegendimage{color1bg,line width = 1pt }
        \addlegendimage{error bar legend={color2bg}}
        \addlegendimage{error bar legend={color5bg}}
        \addlegendimage{error bar legend={color3bg}}
        \addlegendimage{error bar legend={color4bg}}

        \addplot[color1bg,line width = 1 pt] {(x+2)/(x+3)};
        
        \addplot[mark=nomark,color=color2bg, mark size=1pt,dashed, mark options={thin,solid},error bars/.cd, x dir=none, y dir=both, y explicit,
            error bar style={color=color2bg,solid, line width = 1 pt}, error mark options = {line width = 8 pt}] 
            table [x expr = \thisrow{no. of rounds}, y expr =\thisrow{mitigated distillation rate}, y error expr = \thisrow{std}, col sep=comma, mark=square] {./epr_dist_rate_v_rounds_mit_aachen_nobarrier.csv};
        \addplot[mark=nomark,color=color5bg, mark size=1pt,dashed, mark options={thin,solid},error bars/.cd, x dir=none, y dir=both, y explicit,
            error bar style={color=color5bg,solid, line width = 1 pt}, error mark options = {line width = 8 pt}] 
            table [x expr = \thisrow{no. of rounds}, y expr =\thisrow{distillation rate}, y error expr = \thisrow{std}, col sep=comma, mark=square] {./data/epr_dist_rate_v_rounds.csv};
            
        \addplot[mark=nomark,color=color3bg, mark size=1pt,dashed, mark options={thin,solid},error bars/.cd, x dir=none, y dir=both, y explicit,
            error bar style={color=color3bg,solid, line width = 1 pt}, error mark options = {line width = 8 pt}] 
            table [x expr = \thisrow{no. of rounds}, y expr =\thisrow{mitigated distillation rate}, y error expr = \thisrow{std}, col sep=comma, mark=square] {./data/epr_dist_rate_v_rounds_mit.csv};

        \addplot[mark=nomark,color=color4bg, mark size=1pt,dashed, mark options={thin,solid},error bars/.cd, x dir=none, y dir=both, y explicit,
            error bar style={color=color4bg,solid, line width = 1 pt}, error mark options = {line width = 8 pt}] 
            table[x=x,y=y,y error = error]{
            x y error 
            1 0.751 0.03
            };

        \addlegendentry{theoretical success rate}

        \addlegendentry{mitigated experiment data (aachen)}
        \addlegendentry{unmitigated experiment data (quebec)}
        \addlegendentry{mitigated experiment data (quebec)}

        \addlegendentry{\cite{OpticalSagnacRPD}}

        \end{axis}
    \end{tikzpicture}
    \hspace{15pt}
    \caption{Comparison of the success probability of distillation measured on \texttt{ibm\_aachen} and \texttt{ibm\_quebec}. The solid curve shows the ideal theoretical prediction derived in Ref.~\cite{OpticalSagnacRPD}. Experimental points correspond to mitigated data obtained using M3. Error bars denote $1\sigma$ uncertainties estimated from five consecutive experimental trials. Data in the unshaded region correspond to specific party distillation, while the shaded region indicates the random party distillation stage.}
    \label{fig:epr_dist_rate_comparison}
\end{figure}

To compare the two devices, we estimated the success probability of random party distillation for up to four rounds using the analysis procedure described in the main text. Success probabilities were obtained by counting events in which exactly one of Alice, Bob, and Charlie measured ``1'' on the corresponding ancilla qubit over $10^4$ trials. Figure~\ref{fig:epr_dist_rate_comparison} compares the mitigated success probabilities measured on \texttt{ibm\_aachen} and \texttt{ibm\_quebec}.

We find that the performance on \texttt{ibm\_aachen} remains close to the noiseless theoretical prediction over multiple rounds, whereas the success probability on \texttt{ibm\_quebec} begins to saturate after approximately two rounds. This behavior is consistent with the more rapid degradation of the underlying \(W\)-state fidelity on \texttt{ibm\_quebec}, as discussed later in this section. We also find that the estimated rate of success without readout error mitigation on \texttt{ibm\_quebec} will exceed that in which M3 mitigation is employed due to ``false-positive" success events from random bit flips in the ancillary qubits.

\begin{figure}[]
    \centering 

    \begin{tikzpicture}[scale=1]
        \begin{axis}[
            xlabel={No. of Rounds},
            ylabel={Fidelity to \textit{W} State},
            xmin=0, xmax=5,
            ymin=0.3, ymax=1,
            xtick={0,1,2,3,4},
            domain = {0:5},
            ytick={0.3,0.4,0.5,0.6,0.7,0.8,0.9,1},
            legend pos=north east,
            ymajorgrids=true,
            xmajorgrids=true,
            grid style=dashed,
            error bars/y dir=both, 
            error bars/y explicit,  
            width=0.6\textwidth,
            height=0.45\textwidth
        ]   
        
        \addlegendimage{error bar legend={color1bg}}
        \addlegendimage{error bar legend={color2bg}}
        \addlegendimage{error bar legend={color4bg}}

        \addplot[mark=nomark,color=color1bg, mark size=1pt,dashed, mark options={thin,solid},error bars/.cd, x dir=none, y dir=both, y explicit,
        error bar style={color=color1bg,solid, line width = 1 pt}, error mark options = {line width = 8 pt}] 
        table [x expr = \thisrow{no. of rounds}, y expr =\thisrow{w_exp_list_mit}, y error expr = \thisrow{w_std_mit}, col sep=comma, mark=square] {./data/w_state_fid_5.csv};
        
        \addplot[mark=nomark,color=color2bg, mark size=1pt,dashed, mark options={thin,solid},error bars/.cd, x dir=none, y dir=both, y explicit,
            error bar style={color=color2bg,solid, line width = 1 pt}, error mark options = {line width = 8 pt}] 
            table [x expr = \thisrow{no. of rounds}, y expr =\thisrow{w_exp_list_mit}, y error expr = \thisrow{w_std_mit}, col sep=comma, mark=square] {./w_fid_dd_notwirling_opt_qubits_opt2_nobarrier.csv};

        \addplot[mark=nomark,color=color2bg, mark size=1pt,dashed, mark options={thin,solid},error bars/.cd, x dir=none, y dir=both, y explicit,
            error bar style={color=color4bg,solid, line width = 1 pt}, error mark options = {line width = 8 pt}] 
            table[x=x,y=y,y error = error]{
            x y error 
            0 0.866 0.007
            };

        \addlegendentry{\texttt{ibm\_quebec}}

        \addlegendentry{\texttt{ibm\_aachen}}

        \addlegendentry{\cite{OpticalSagnacRPD}}


        \end{axis}
    \end{tikzpicture}
    \hspace{15pt}
    \caption{The evolution of \textit{W} state fidelity on \texttt{ibm\_quebec} (blue) and \texttt{ibm\_aachen} (orange), determined by a series of measurements in the Pauli basis. Error bars corresponding to 1-$\sigma$ uncertainty were obtained by repeating experiments over five consecutive trials. The initial \textit{W} state fidelity of \cite{OpticalSagnacRPD} is shown in red for comparison.}
    \label{fig:w_state_fidelity_evolution_comparison}
\end{figure}
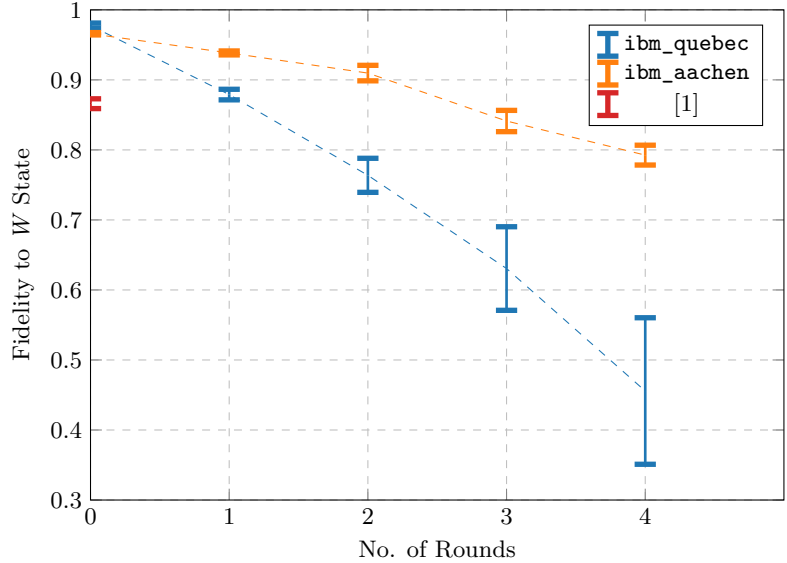

To quantify the degradation of the resource state across rounds, we estimated the expectation value of the three-qubit \(W\)-state projector at the beginning of each round of protocol execution, following the measurement procedure described in the main text. The resulting fidelities are shown in Fig.~\ref{fig:w_state_fidelity_evolution_comparison}.

The \(W\)-state fidelity on \texttt{ibm\_quebec} decreases substantially faster with round number than on \texttt{ibm\_aachen}. A linear fit to the measured data yields an approximate decay rate of $0.13$ per round on \texttt{ibm\_quebec}, compared with $0.04$ per round on \texttt{ibm\_aachen}. This trend supports the interpretation that the reduced multi-round success probability on \texttt{ibm\_quebec} is closely related to a more rapid deterioration of the intermediate \(W\)-state.

\begin{figure}[H]
    \centering
    \includegraphics[width=0.9\linewidth]{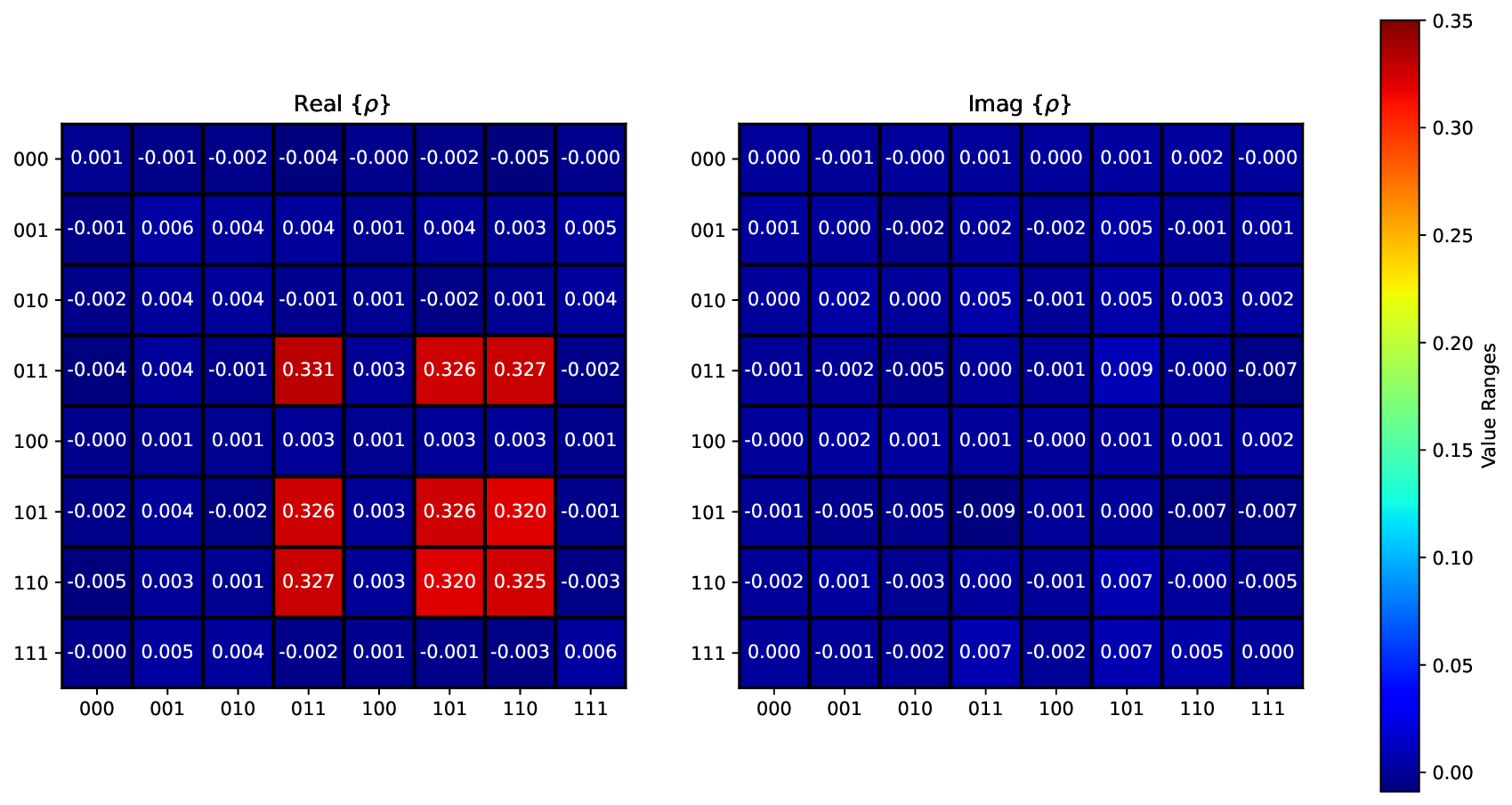}
    \caption{The density matrix of a \textit{W} state created on \texttt{ibm\_quebec}. The fidelity to the ideal scenario is found to be $0.958 \pm 0.007$ (1-$\sigma$ error).}
    \label{fig:w_state}
\end{figure}

One could also estimate the fidelity and other properties of a system from its density matrix, estimated by a process of quantum state tomography. Fig. ~\ref{fig:w_state} shows the reconstructed density matrix of the initial \(W\) state, obtained using three-qubit quantum state tomography prior to protocol execution on \texttt{ibm\_quebec}. We note that the estimated density matrix obtained on \texttt{ibm\_aachen} has been omitted from this study due to the fact that the results on the two processors appear nearly identical after \textit{W} state initialization. The similarity in initial \textit{W} state fidelity between \texttt{ibm\_quebec} and \texttt{ibm\_aachen} further supports our claim that the two processors differ only during protocol execution. Readout errors were mitigated using the tensor-product noise model implemented in the Qiskit Experiments framework~\cite{PhysRevA.103.042605,Kanazawa_Qiskit_Experiments_A_2023}. In this approach, the measurement errors are assumed to be predominantly local, such that a calibration matrix obtained from independently prepared basis states can be used to map observed outcomes to mitigated probabilities. From the mitigated data, we estimate the fidelity with the ideal \(W\) state to be $0.958 \pm 0.007$, where the uncertainty denotes a $1\sigma$ statistical error.

\section{Measuring Conditional Observables}

In this section, we list the experimentally estimated probabilities used to evaluate the expected entanglement reported in the main text. For a protocol of fixed depth, \(P(W)\) denotes the probability that the protocol continues to the next round, while \(P(\mathrm{EPR})\) denotes the probability of successful extraction of an EPR pair at that round. The entries labelled ``weak measurement'' correspond to random party distillation rounds, whereas the entries labelled ``strong measurement'' correspond to the final specific-party distillation step. All values reported below are obtained from mitigated measurement data (using M3 measurement error mitigation; see main text for more details). Table \ref{tab:empirical_probabilities0} shows the success probability and expected entanglement for a specific party distillation protocol. All probabilities of successful distillation and lower bounds for entanglement of formation used to calculate the expected entanglement $\expval{E}$ can be found in tables \ref{tab:empirical_probabilities1} - \ref{tab:empirical_probabilities4} and \ref{tab:eof_strong_measurement}, respectively. All experimental data presented in this section was obtained while using the device \texttt{ibm\_aachen}.

\begin{table}[H]
    \centering
    \begin{tabular}{p{5em}p{8em}p{16em}p{16em}} \hline \hline 
         \textbf{Round}& \textbf{P(W)}&  \textbf{P(EPR) (weak measurement)}& \textbf{P(EPR) (strong measurement)}\\ \hline 
         0 & - & - & 0.6643\\ \hline
    \end{tabular}
    \caption{The estimated probabilities corresponding to a W state and EPR pair being extracted from round $i$ of a \textit{specific} party distillation protocol. The expected entanglement $\expval{E}$ for a single-round random distillation protocol on \texttt{ibm\_aachen} is $\expval{E}= 0.5972$.} 
    \label{tab:empirical_probabilities0}
\end{table}

\begin{table}[H]
    \centering
    \begin{tabular}{p{5em}p{8em}p{16em}p{16em}} \hline \hline 
         \textbf{Round}& \textbf{P(W)}&  \textbf{P(EPR) (weak measurement)}& \textbf{P(EPR) (strong measurement)}\\ \hline 
         0 & - & - & 0\\ 
         1 & 0.6291 & 0.3270 & 0.4246 \\ \hline
    \end{tabular}
    \caption{The estimated probabilities corresponding to a W state and EPR pair being extracted from round $i$ for a single round protocol (with strong measurement). The expected entanglement $\expval{E}$ for a single-round random distillation protocol on \texttt{ibm\_aachen} is $\expval{E}= 0.6379$}. 
    \label{tab:empirical_probabilities1}
\end{table}

\begin{table}[H]
    \centering
    \begin{tabular}{p{5em}p{8em}p{16em}p{16em}} \hline \hline 
         \textbf{Round}& \textbf{P(W)}&  \textbf{P(EPR) (weak measurement)}& \textbf{P(EPR) (strong measurement)}\\ \hline 
         0 & - & - & 0\\ 
         1 & 0.6815 & 0.2843 & 0\\ 
         2 & 0.4271 & 0.2235 & 0.2868 \\\hline
    \end{tabular}
    \caption{The estimated probabilities corresponding to a W state and EPR pair being extracted from round $i$ for a two-round protocol (with strong measurement). The expected entanglement $\expval{E}$ for a two-round random distillation protocol on \texttt{ibm\_aachen} is $\expval{E}= 0.5996$}.
    \label{tab:empirical_probabilities2}
\end{table}

\begin{table}[H]
    \centering
    \begin{tabular}{p{5em}p{8em}p{16em}p{16em}} \hline \hline 
         \textbf{Round}& \textbf{P(W)}&  \textbf{P(EPR) (weak measurement)}& \textbf{P(EPR) (strong measurement)}\\ \hline 
         0 & - & - & 0\\ 
         1 & 0.7122 & 0.2607 & 0\\ 
         2 & 0.4968 & 0.1948 & 0\\   
         3 & 0.3267 & 0.1500 & 0.2099\\ \hline
    \end{tabular}
    \caption{The estimated probabilities corresponding to a W state and EPR pair being extracted from round $i$ for a three-round protocol (with strong measurement). The expected entanglement $\expval{E}$ for a three-round random distillation protocol on \texttt{ibm\_aachen} is $\expval{E}= 0.5664$}.
    \label{tab:empirical_probabilities3}
\end{table}

\begin{table}[H]
    \centering
    \begin{tabular}{p{5em}p{8em}p{16em}p{16em}} \hline \hline 
         \textbf{Round}& \textbf{P(W)}&  \textbf{P(EPR) (weak measurement)}& \textbf{P(EPR) (strong measurement)}\\ \hline 
         0 & - & - & 0 \\ 
         1 & 0.7508 & 0.2297 & 0\\ 
         2 & 0.5516 & 0.1818 & 0\\   
         3 & 0.3921 & 0.1451 & 0\\  
         4 & 0.2495 & 0.1260 & 0.1624 \\ \hline
    \end{tabular}
    \caption{The estimated probabilities corresponding to a W state and EPR pair being extracted from round $i$ for a four-round protocol (with strong measurement). The expected entanglement $\expval{E}$ for a four-round random distillation protocol on \texttt{ibm\_aachen} is $\expval{E}= 0.5404$}.
    \label{tab:empirical_probabilities4}
\end{table}

\begin{table}[H]
    \centering
    \begin{tabular}{p{14em}p{14em}} \hline \hline 
         \textbf{Round}& \textbf{$\frac{1}{|\mathcal{S}|}\sum_iE(\rho_i^k)$}, \textbf{$\frac{1}{|\mathcal{S}|}\sum_iE(\rho_i^s)$}\\ \hline 
         0 & $0.8990 \pm 0.0036$ \\ 
         1 & $0.8487 \pm 0.0085$ \\ 
         2 & $0.7022 \pm 0.0331$ \\   
         3 & $0.5788 \pm 0.0592$ \\  
         4 & $0.4641 \pm 0.0652$ \\ 
         5 & $0.3214 \pm 0.0474$ \\ 
         6 & $0.2258 \pm 0.1259$ \\ \hline
    \end{tabular}
    \caption{The estimated lower bounds in entanglement of formation $E(\rho_j)$ for each EPR pair extracted at round $i$, using the average mitigated values obtained from Monte Carlo propagation.}
    \label{tab:eof_strong_measurement}
\end{table}

\section{Coherent Deferred-Measurement Implementation of Random Party Distillation}

In the main text, we identify mid-circuit measurements as a significant source of error in our implementation of random party distillation on current superconducting hardware. Motivated by the dephasing, crosstalk, and idling errors associated with mid-circuit readout, we consider here a coherent deferred-measurement implementation of the protocol. In this construction, each weak-measurement operation is replaced by a unitary interaction between the system qubit and an ancillary probe qubit, with the measurement record stored coherently and read out only at the end of the circuit. This follows from the standard principle of deferred measurement and may also be viewed as a Stinespring-type dilation of the corresponding measurement channel (see Ref. \cite{kretschmann2008information}, for example). The cost of this replacement is an increased number of ancilla qubits and additional two-qubit gates.



The revised version of the protocol takes a form similar to that proposed in the main text. Alice, Bob, and Charlie prepare a state $\ket{W}$ of the form

\begin{equation}
    \ket{W} = \frac{1}{\sqrt{3}}\left(\ket{011} + \ket{101} + \ket{110}\right).
\end{equation}

Each ``round" of the protocol once again consists of Alice, Bob, and Charlie weakly measuring their respective qubits. ``Weak measurements" in this case are replaced by weakly coupling an ancillary qubit (probe) to a qubit belonging to the original \textit{W} state (system) using a $CRY(\theta)$ gate. Once again, we define the $CRY(\theta)$ gate:

\begin{align}
    CRY(\theta)_{A_i,i} &= \mathds{1}\otimes\ket{0}\!\!\bra{0} + RY(\theta)\otimes \ket{1}\!\!\bra{1} \nonumber \\
    &= \begin{bmatrix}
        1 & 0 & 0 & 0 \\
        0 & \cos(\frac{\theta}{2}) & 0 & -\sin(\frac{\theta}{2}) \\
        0 & 0 & 1 & 0 \\
        0 & \sin(\frac{\theta}{2}) & 0 & \cos(\frac{\theta}{2})
    \end{bmatrix}.
\end{align}

The choice of $\theta_D$ is related to the strength of the measurement $\epsilon_D$ using the following formula: 

\begin{equation}
    \theta_D  = 2\arcsin\left(\epsilon_D\right).
    \label{eq:weakness_consts}
\end{equation}

The measurement strength $\epsilon_D$ depends on the number of rounds remaining $D$ and is given by 

\begin{equation}
    \epsilon_D = \frac{1}{\sqrt{D+5}}.
    \label{eq:opt_epsilon}
\end{equation}

Previously, this would have been followed by Alice, Bob and Charlie performing projective (mid-circuit) measurements defined by the operators $F = \ket{0}_{A_i}\bra{0}_{A_i}$ and $G = \ket{1}_{A_i} \bra{1}_{A_i}$. In this revised version of the protocol, the three parties perform a series of cascaded CNOT operations between ancillary qubits. In operator form, Alice, Bob, and Charlie perform the following operation: 

\begin{equation}
    \prod_i^D U_{CNOT}^{k(D-i)(D-i+1)} \rho_{wm} \prod_i^D \left(U_{CNOT}^{k(D-i)(D-i+1)}\right)^\dagger,
\end{equation}

where $U_{CNOT}^{kij}$ represent the CNOT gates performed by the $kth$ party (Alice, Bob, or Charlie) between qubits $i$ and $j$:

\begin{equation}
    U_{CNOT}^{kij} = \ket{0}\!\!\bra{0}_{ki}\otimes\ket{0}\!\!\bra{0}_{kj} + \ket{1}\!\!\bra{1}_{ki}\otimes\ket{1}\!\!\bra{0}_{kj} + \ket{1}\!\!\bra{1}_{ki}\otimes\ket{0}\!\!\bra{1}_{kj}
    + \ket{0}\!\!\bra{0}_{ki}\otimes\ket{1}\!\!\bra{1}_{kj}.
\end{equation}

\begin{figure}[H]
    \centering
    \includegraphics[width=1\linewidth]{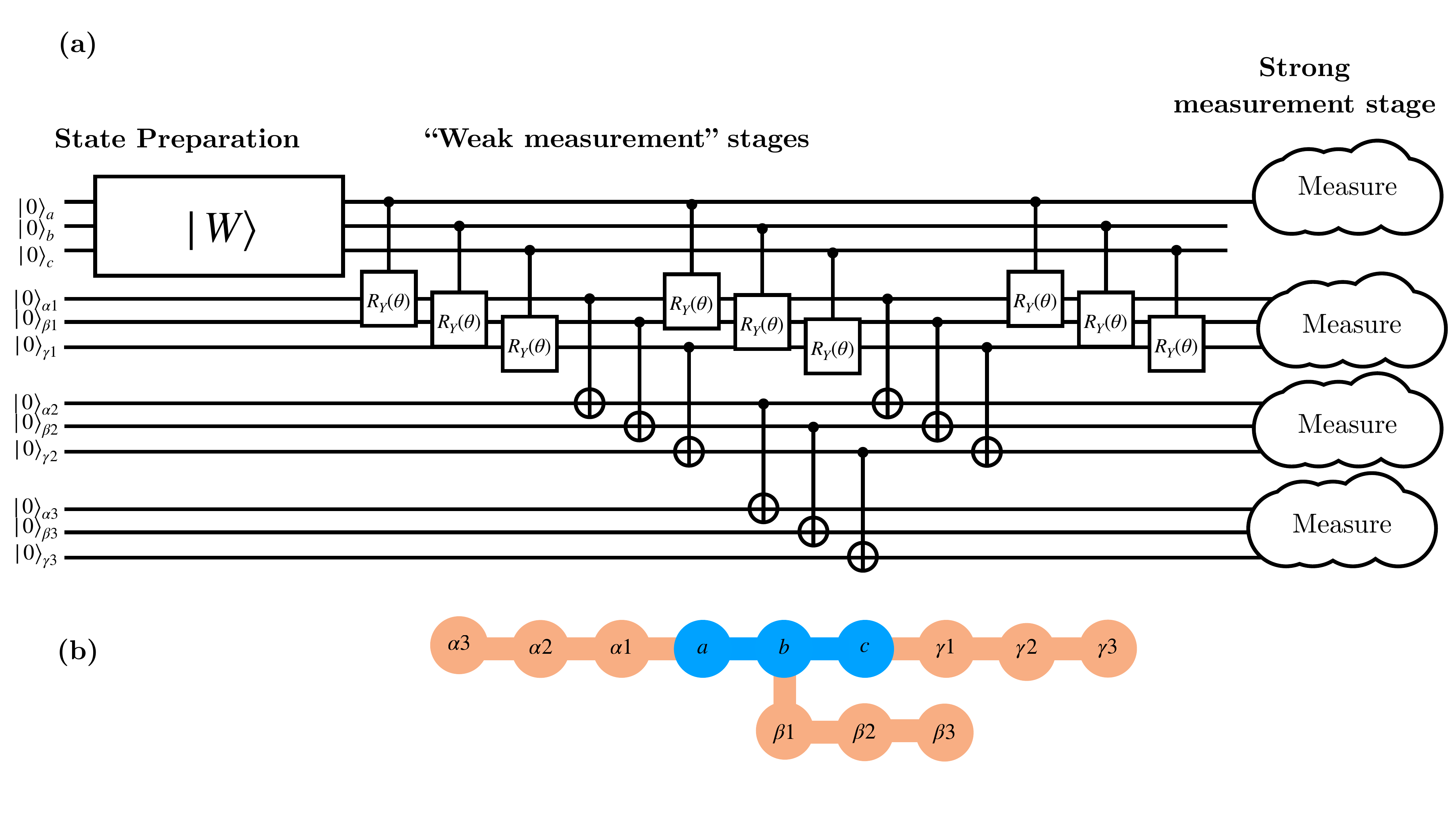}
    \caption{(a) A circuit diagram showing our proposed random party distillation protocol \textit{without} need for mid-circuit measurements. (b) The corresponding qubit connectivty graph showing signal qubits (blue) and ancillae (orange). Mid-circuit measurements are replaced with an additional step where information carried by qubits (i.e. weak measurement outcomes) is transferred to nearest, next-nearest, etc. neighbours so that $\alpha1,\beta1,\gamma1$ may be reused for subsequent rounds.}
    \label{fig:random_party_no_mcm}
\end{figure}

The state $\rho_{wm}$ represents the \textit{W} state possessed by the three parties, \textit{after} having performed the controlled rotations $CRY(\theta)$:

\begin{equation}
    \rho_{wm} = CRY(\theta)_{\alpha1,a}CRY(\theta)_{\beta1,b}CRY(\theta)_{\gamma1,c}\ket{W}\!\!\bra{W}CRY(\theta)_{\alpha1,a}^\dagger CRY(\theta)_{\beta1,b}^\dagger CRY(\theta)_{\gamma1,c}^\dagger,
\end{equation}

where we have assumed that the ancillary qubits ($\alpha_1,\beta_1,\gamma_1$) are initialized in the ground state. Fig. \ref{fig:random_party_no_mcm} shows a circuit diagram and connectivity graph of the proposed protocol. 

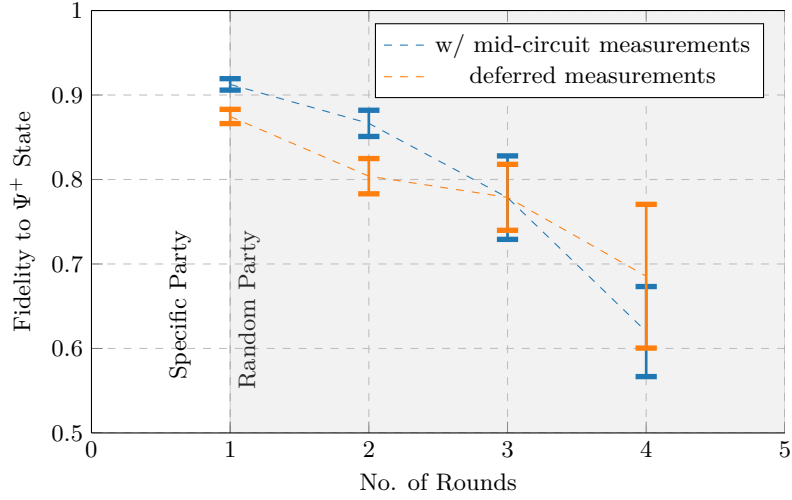
\begin{figure}[H]
    \centering 

    \begin{tikzpicture}[scale=1]
        \begin{axis}[
            xlabel={No. of Rounds},
            ylabel={Fidelity to $\Psi^+$ State},
            xmin=0, xmax=5,
            ymin=0.5, ymax=1,
            xtick={0,1,2,3,4,5},
            domain = {0:7},
            ytick={0.5,0.6,0.7,0.8,0.9,1},
            legend pos=north east,
            ymajorgrids=true,
            xmajorgrids=true,
            grid style=dashed,
            axis y line* = left,
            error bars/y dir=both, 
            error bars/y explicit,  
            width=0.6\textwidth,
            height=0.4\textwidth
        ]   
        \node at (axis cs:0.5,0.75) [anchor=north east,rotate=90] {Specific Party};
        \node at (axis cs:1,0.75) [anchor=north east,rotate=90] {Random Party};
        \filldraw[fill=lightgray,opacity=0.2] (100,0) rectangle (500,500);
        
        \addplot[mark=nomark,color=color1bg, mark size=1pt,dashed, mark options={thin,solid},error bars/.cd, x dir=none, y dir=both, y explicit,
            error bar style={color=color1bg,solid, line width = 1 pt}, error mark options = {line width = 8 pt}] 
            table [x expr = \thisrow{no. of rounds}, y expr =\thisrow{w_exp_list_mit}, y error expr = \thisrow{w_std_mit}, col sep=comma, mark=square] {./data/epr_fid_no_mcm.csv}; \label{fidelity_no_mcm}

        \addplot[mark=nomark,color=color2bg, mark size=1pt,dashed, mark options={thin,solid},error bars/.cd, x dir=none, y dir=both, y explicit,
            error bar style={color=color2bg,solid, line width = 1 pt}, error mark options = {line width = 8 pt}] 
            table [x expr = \thisrow{no. of rounds}, y expr =\thisrow{w_exp_list_mit}, y error expr = \thisrow{w_std_mit}, col sep=comma, mark=square] {./data/epr_fid_w_mcm.csv}; \label{fidelity_w_mcm}

        \addlegendimage{error bar legend={color1bg}}
        \addlegendimage{error bar legend={color2bg}}
        
        \addlegendentry{w/ mid-circuit measurements}
        \addlegendentry{deferred measurements}
        \end{axis}
          
    \end{tikzpicture}
    \hspace{15pt}
    \caption{The evolution of EPR state fidelity on \texttt{ibm\_fez}, measured by a projection onto the $\Psi^+$ state in the Pauli basis. Error bars corresponding to 1-$\sigma$ uncertainty were obtained by repeating experiments over five consecutive trials. We observe a comparable evolution in the distilled EPR state fidelity using both mid-circuit- and deferred-measurement schemes, largely due to the ``overhead" added by additional entangling gate operations in the deferred measurement scheme.}
    \label{fig:epr_fid_comp_mcm}
\end{figure}

Following this proposal, we demonstrated the protocol's performance on the 156 qubit Heron r2 processor, \texttt{ibm\_fez}. The evolution in distilled pair fidelity for both the protocol with and without mid-circuit measurements is shown in Fig. \ref{fig:epr_fid_comp_mcm}. Similarly to the experiments presented in the body-text, XY4 dynamical decoupling sequences were used, followed by M3 for measurement-error mitigation.  

As shown in Fig. \ref{fig:epr_fid_comp_mcm}, we find that the state of the system after a successful distillation event has fidelity comparable to the EPR state, regardless of whether or not mid-circuit measurements are used. The resulting fidelity is unsurprising, considering the number of CNOT operations required for protocol execution, introducing both gate error (median $2.73\times10^{-3}$) and cumulative gate durations comparable to the process of mid-circuit measurement ($68$ ns/ECR operation). Much like our protocol requiring mid-circuit measurements, we predict that further improvements will arise from shorter gate operations to prevent dephasing from the external environment. 
\section{Simulating Noisy Random Party Distillation}
\label{sec:simulation}

To interpret the experimentally observed degradation in protocol performance across rounds, we consider a simplified noise model in which the three-qubit resource state is replaced at each stage by a white-noise mixture of the form
\begin{equation}
    \rho_W^{\mathrm{imp}} = (1-p)\ket{W}\!\bra{W} + \frac{p}{8}\mathds{1},
    \label{eq:noisy_w}
\end{equation}
where \(\ket{W}\) denotes the ideal three-qubit \(W\) state. This model is not intended to reproduce the full device-level noise, but rather to capture, at a coarse-grained level, the effect of round-dependent deterioration in the resource-state fidelity.

We first evaluate a single round of random party distillation assuming perfect gate operations but an imperfect initial \(W\) state of the form in Eq.~\eqref{eq:noisy_w}. The success probability is obtained from the expectation values of the permutation-symmetrized projectors \(\mathrm{Per}\{F\otimes F\otimes G\}\), where \(F=\ket{0}\!\bra{0}\) and \(G=\ket{1}\!\bra{1}\) act on the ancilla qubits. Conditional on a successful event, the entanglement of the resulting two-qubit state is quantified through its concurrence 
\begin{equation}
    \mathcal{C}(\rho) = \max\{0,\lambda_1 - \lambda_2-\lambda_3 - \lambda_4\},
\end{equation}
where $\lambda_i$s are the eigenvalues (in decreasing order) of the Hermitian matrix $R$
\begin{equation}
    R = \sqrt{\sqrt{\rho}\tilde{\rho}\sqrt{\rho}},
\end{equation}
\begin{equation}
    \tilde{\rho} = (\sigma_y\otimes\sigma_y )\rho^* (\sigma_y \otimes \sigma_y),
\end{equation}
and the associated entanglement of formation
\begin{equation}
    E(\mathcal{C}) = h\left(\frac{1 + \sqrt{1-\mathcal{C}^2}}{2}\right),
\end{equation}
where $h(\cdot)$ is the binary entropy function (described in the main text).

\begin{figure}[H]
    \centering

    \begin{tikzpicture}[node distance=3cm]
        \node (start) [startstop] {Prepare  an imperfect \textit{W} state in simulation $\rho_W = (1-p)\ket{W}\!\!\bra{W} + \frac{p}{8}\mathds{1}$};
        \node (random_party) [process, below of=start] {Perform one round of random party distillation, using perfect gate operations.};

        \node (assignment1) [thickprocess, below of=random_party] {Record probability of distillation and entanglement of formation $E$ of distilled pair.};

        \node (assignment2) [thickprocess, below of=assignment1] {Prepare imperfect \textit{W} state with fidelity \textit{lower} than that of previous stage.};
        
        \node (stop) [startstop, below of=assignment2] {Perform specfic party distillation, record probability of success and $E$.};
;

        \draw [arrow] (start) -- (random_party);
        \draw [arrow] (random_party) -- (assignment1);
        \draw [arrow] (assignment1) -- (assignment2);
        \draw [arrow] (assignment2) --node[anchor=east] {Continue after $N$ rounds} (stop);
        
        \path[]
            (assignment2) edge[thick, ->,bend left=90,>=stealth] node [left] {Repeat for $N$ rounds} (random_party);
    \end{tikzpicture}

    \caption{A flow-chart depicting the process by which the success rate of a one-way distillation protocol is obtained from estimating the expectation values of Pauli observables.}
    \label{fig:flow_chart}
\end{figure}
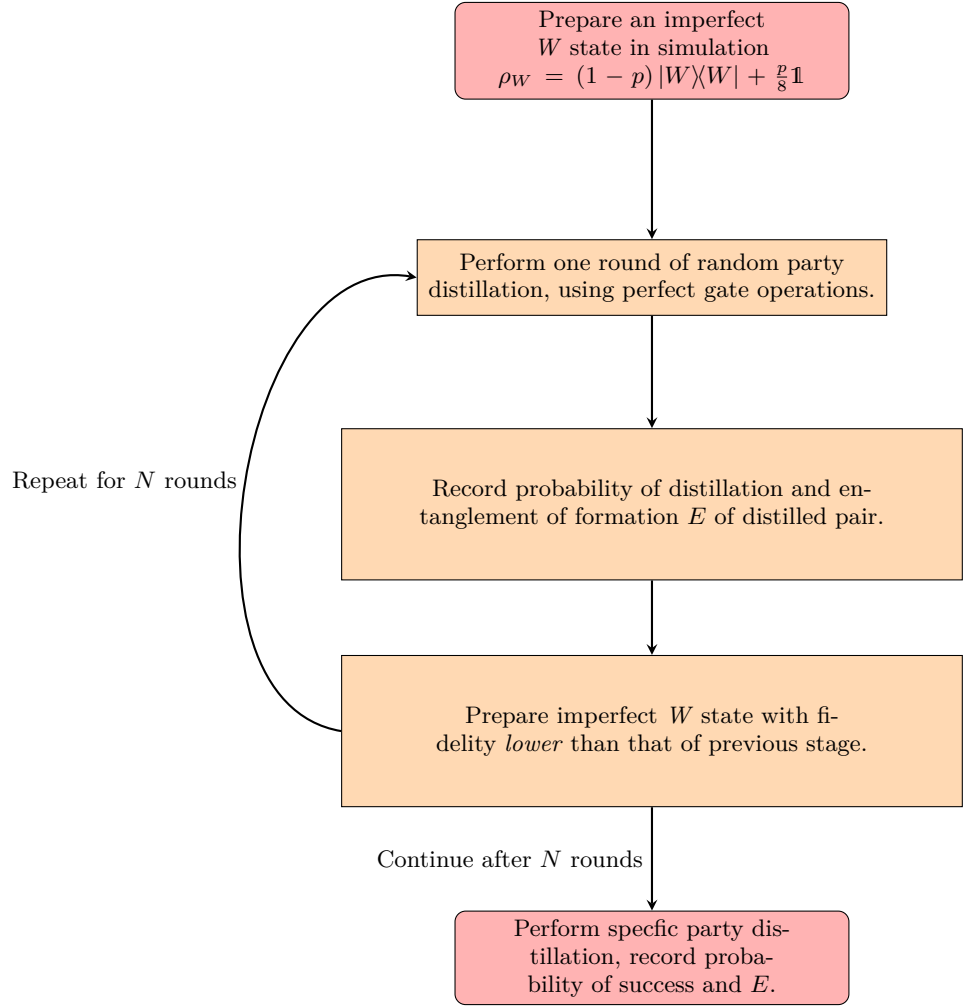

To model multi-round operation, we repeat this procedure while reducing the \(W\)-state fidelity from one round to the next, thereby mimicking the experimentally observed deterioration of the intermediate resource state. After the random-distillation rounds are completed, a final specific-party distillation step is applied. A schematic summary of the simulation procedure is shown in Fig.~\ref{fig:flow_chart}.

\begin{figure}[H]
    \centering
    \includegraphics[width=0.7\linewidth]{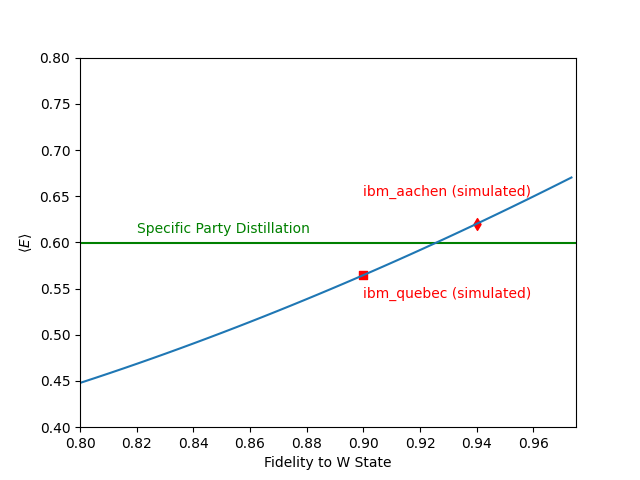}
    \caption{The Fidelity of a noisy \textit{W} state $\rho_W$ to the ideal \textit{W} state \textit{after} 1 round of random party distillation (blue) when compared against specific party distillation (green) starting with fidelity of 0.97. We find that the fidelity possessed by \texttt{ibm\_aachen} (labeled by the red diamond) after a \textit{single} round of random party distillation meets the threshold at which the protocol's expected entanglement $\expval{E}$ exceeds that of specific party distillation. In contrast, the \textit{W} state fidelity after a single round of random party distillation on \texttt{ibm\_quebec} is insufficient to surpass the expected entanglement of specific party distillation.}
    \label{fig:rand_v_spec_dist}
\end{figure}

As an example, we simulate the protocol's performance for a single round of random party distillation followed by specific party distillation. Fig. \ref{fig:rand_v_spec_dist} displays the expected entanglement $\langle E\rangle$ obtained from a single round of random party distillation for fidelity values (in the range $F\in [0,0.97]$) of the noisy $W$ state after the first round of weak measurements.

As expected, we find that an improvement in $\expval{E}$ occurs when the fidelity of the noisy \textit{W} state exceeds $\sim0.93$. Furthermore, we find the model's prediction of $\expval{E} \approx 0.555$ for a first-round fidelity of $F\approx 0.9$ and $F\approx 0.94$ tracks our experimental results obtained using both \texttt{ibm\_quebec} and \texttt{ibm\_aachen}, respectively, with high accuracy (within 4.3\% and 2.8\%). 

Indeed, the close agreement between our simplistic model using perfect gate operations and experiment supports the claim that mid-circuit measurements are the dominant source of error in our systems (rather than gate errors). 

\bibliography{random_party.bib}